\documentclass[fleqn,usenatbib]{mnras}

\usepackage{newtxtext,newtxmath}

\usepackage[T1]{fontenc}
\usepackage{orcidlink}
\usepackage{tabularx}
\usepackage{ulem}
\DeclareRobustCommand{\VAN}[3]{#2}
\let\VANthebibliography\thebibliography
\def\thebibliography{\DeclareRobustCommand{\VAN}[3]{##3}\VANthebibliography}

\usepackage{graphicx}	
\usepackage{amsmath}	
\usepackage{balance}
\usepackage{subcaption}  

\usepackage{pdflscape}
\usepackage{lipsum}

\usepackage{longtable}
\usepackage{booktabs}
\usepackage{multirow}

\usepackage{mathtools}
\newcommand{\defeq}{\vcentcolon=}

\graphicspath{{./images/}}
\newcommand*\diff{\mathop{}\!\mathrm{d}}

\title[Spinning dust spectral signatures]{Spectral Signatures of Spinning Dust from Grain Ensembles in Diverse Environments: A Combined Theoretical and Observational Study}

\author[Z. Zhang et al.]{
Zheng Zhang$^{1}$\orcidlink{0000-0002-9154-2803}\thanks{E-mail: zheng.zhang@manchester.ac.uk (ZZ)}, Jens Chluba$^{1}$, 
Roke Cepeda-Arroita$^{2,3}$, and Jos\'{e} Alberto Rubi\~no-Mart\'{i}n$^{2,3}$
\\
$^{1}$Jodrell Bank Centre for Astrophysics, University of Manchester, Manchester, M13 9PL, UK \\
$^{2}$Instituto de Astrof\'{i}sica de Canarias, 38200 La Laguna, Tenerife, Canary Islands, Spain \\
$^{3}$Departamento de Astrof\'{i}sica, Universidad de La Laguna (ULL), 38206 La Laguna, Tenerife, Spain
}

\date{Accepted XXX. Received YYY; in original form ZZZ}

\pubyear{\the\year{}}

\begin{document}
\label{firstpage}
\pagerange{\pageref{firstpage}--\pageref{lastpage}}
\maketitle

\begin{abstract}
Recent observations of anomalous microwave emission (AME) reveal spectral features that are not readily reproduced by spinning dust models.
We examine how dust grain distributions and environmental parameters determine the peak frequency and spectral width of AME spectral energy distribution (SED). 
Using Monte Carlo sampling and global sensitivity analysis, we find that AME features are dominantly controlled by three parameters: grain size, shape, and a phase-dependent environmental parameter.
We also quantify the effects of SED broadening from ensembles of these dominant parameters, finding that the level of tension with observations is strongly phase dependent:
Molecular Cloud (MC) is fully consistent,  Dark Cloud (DC) shows minor deviations, and H\textsc{ii} regions exhibit significant offsets in peak frequency. 
The discrepancy in H\textsc{ii} echoes the observed depletion of small dust grains, particularly polycyclic aromatic hydrocarbons (PAHs), in H\textsc{ii} regions.
However, an observational H\textsc{ii} region may still have AME originating from nearby non-H\textsc{ii} clouds. In this case, model calculations for H\textsc{ii} regions would be inappropriate.
Reproducing MC and DC AME features requires ensemble variations in both grain size and environmental parameters are required to reproduce the observed spread in peak frequency and spectral width.
We further propose moment expansion and emulation-based inference methods for future AME spectral analysis.
\end{abstract}



\section{Introduction}
\label{sec:intro}

Interstellar dust, though comprising only about 1\% of the total interstellar medium mass, exerts a profound influence on a wide range of astrophysical processes. 
Dust grains are essential for understanding galactic evolution, star and planet formation, and radiative transfer processes through the interstellar medium (ISM). Furthermore, they shape the observed appearance of galaxies across the electromagnetic spectrum,  from microwaves to ultraviolet light, and thus play a pivotal role in observations \citep{draine2003interstellar}.

A source of both astrophysical interest and concern for cosmological surveys is the spinning electric dipole emission from interstellar dust grains \citep{DL98b}.
This emission is widely recognised as the primary source of anomalous microwave emission (AME; see, for example, \cite{dickinson2018state} for a recent review).
At low radio frequencies, the emission is dominated by very small dust grains, which can attain rapid rotation through a combination of collisional and radiative torques, as well as internal processes such as vibrational-rotational coupling \citep{hoang2010}.
The resulting spectral energy density (SED) depends on the rotational distribution and electric dipole moments of the dust grains, which are determined by their intrinsic properties and the ambient ISM conditions.

Over the past few decades, substantial progress has been made in modelling the spinning dust emission mechanism  \citep{DL98b, AHD09, hoang2010, hoang2011, SAH11, hoang2016spinning, hoang2016unified, draine2016quantum, hensley2017modeling, zhang2025spydust}. 
However, the high complexity of the interstellar environment makes precision modelling extremely challenging. 
Theoretical efforts focus on calculating spinning dust emission conditional on the properties of dust grains and interstellar medium.
Current simulators of spinning dust emission can further account for an assumed distribution of grain sizes and shapes within a specified ISM environment.
However, observed spectra are likely to result from environments with internal variations or distinct ISM phases and grain distributions that differ from those assumed in the models.

On the other hand, attempts to fully constrain spinning dust models directly from AME observations are hindered by the curse of dimensionality, with parameter degeneracies and a high computational cost of exploring such a large parameter space with numerical simulations. 
As a result, most observational studies have instead focused on extracting characteristic AME features -- including the peak frequency and spectral width -- and interpreting them using state-of-the-art simulation tools [e.g. \textsc{SpDust2} \citep{SAH11} and \textsc{SpyDust} \citep{zhang2025spydust}].
Notably, \cite{cepedaarroita2025} recently analysed the spectral properties of AME in $144$ Galactic clouds by combining low-frequency maps from S-PASS \citep{spass_release}, C-BASS (Taylor et al. in prep.), and QUIJOTE \citep{mfiwidesurvey} with ancillary maps.
The observed AME spectra were fitted using a log-Gaussian model to extract the spectral signatures, including the peak and width. 
The analysis shows that the peak frequencies are consistently higher and the SED widths consistently broader than predicted by standard \textsc{SpDust2} and \textsc{SpyDust} simulations.
In particular, the observed SED widths are typically $\simeq 0.6$\footnote{See the SED model in eq~(\ref{eq: sed fit model}) for the meaning of these values.}, compared to $\simeq 0.4$ in the simulations. Peak frequency discrepancies are particularly evident in the idealised RN and PDR cases.
\cite{fernandez2023quijote} further demonstrated that these discrepancies are not limited to individual objects; the diffuse AME throughout the Galactic plane exhibits systematically high median width values ($\simeq 0.56$) that remain in tension with current theoretical models.

This discrepancy between observations and simulations requires detailed analysis. 
Given the hybrid nature of the simulators -- combining deterministic modelling with assumed environmental conditions and grain distributions -- the discrepancy should be understood from two complementary perspectives: (i) limitations of the theoretical model, which involves refining our understanding of the astrophysical environment and rotational dynamics, and (2) revisiting the assumptions of dust grain and environment models on which these simulations are based.

In this paper, we hold fixed the underlying theoretical model implemented in \textsc{SpyDust} and explore a general joint distribution of grain size, shape, and environmental conditions.
Before presenting the details of our analysis, we first explain why this regime is of particular interest:
(1) \textit{It remains relatively unexplored.} 
Although the rotational distribution of spinning dust grains has been studied extensively, the roles of grain size, shape, and environment, in particularly their ensemble effects, have received less attention in the context of AME theory, despite potentially having a significant impact on the predicted AME spectrum. 
For example, \citet{zhang2025spydust} demonstrated that grain oblateness, within a single-size population, can substantially modify the shape of the spinning dust SED when the same rotational statistics framework is applied.
(2) \textit{Current prescriptions for grain size and shape should not be regarded as definitive.} 
Under the assumption of a carbonaceous composition consistent with polycyclic aromatic hydrocarbons (PAHs), the commonly adopted size distributions for spinning dust grains were originally derived by fitting diffuse infrared emission \citep{LD01a, WD01}, which may probe a different grain population. 

However, in addition to the aforementioned observational discrepancies, there are several other reasons to question these assumptions:
(1) PAHs are not the only viable candidates for spinning dust emission [see e.g. \cite{chuss2022tracing}]. 
Other nanocarbonaceous grains may also contribute to mid-infrared continuum emission or specific spectral features \citep{li2003interaction, jones2013evolution}, which could potentially bias abundance estimates, even though none of these species alone can reproduce the full range of PAH emission characteristics \citep{iglesias2005electric}. 
Furthermore, \cite{hensley2017modeling} proposed nanosilicates as an alternative AME carrier whose infrared signatures would differ substantially and potentially appear only as broad, weak features. 
Taken together, these considerations imply that the populations of grains responsible for infrared and spinning dust emissions may not be identical.
(2) Both infrared emission and the extinction curve have been observed to vary across environments \citep{fitzpatrick1999correcting}, implying that size distributions inferred from a case of diffuse emission may not be universally applicable to all ISM phases.
(3) Interstellar dust is inherently heterogeneous, and its properties -- including
size distribution, chemical composition, and internal structure -- are not static but dynamically
evolve in response to local physical conditions such as radiation field intensity and hardness,
gas density, and gas dynamics.
In short, any rapidly rotating grain with an electric dipole can produce spinning dust emissions. Therefore, focusing exclusively on PAHs requires strong justification, while
observational evidence for a direct association between AME and PAHs remains mixed: Large-scale studies find that AME correlates more strongly with total dust emission than with PAH abundance \citep{hensley2016case,sponseller2025statistical}, although analyses of individual regions have reported stronger correlations with PAH-related emission \citep{bell2019investigation}. These results suggest that while PAHs remain plausible carriers, the observational evidence does not uniquely favor them, and other ultrasmall grain populations may also contribute to AME.

Motivated by these considerations, we investigate how variations in grain size, shape, and environmental conditions affect the spectral features of spinning dust emission, with particular emphasis on the peak frequency and spectral width.
All simulations are implemented using \textsc{SpyDust}, which retains the rotational statistics of \textsc{SpDust2} but allows greater flexibility in modelling arbitrary grain geometries \citep{zhang2025spydust}.

This study is structured around three hierarchical questions. First, in the high-dimensional parameter space defined by ISM environmental and grain properties, which parameters most strongly control the SED shape? Identifying these dominant drivers allows us to reduce the effective dimensionality of the problem and render a systematic exploration of the theory tractable.

Second, assuming a general distribution for the key parameters, are the observed AME features consistent with theoretical expectations? To address this, we compare the model-allowed SED features, generated via Monte Carlo sampling, with recent AME feature catalogues from \citet{cepedaarroita2025} for three representative ISM environments -- Molecular Cloud (MC), Dark Cloud (DC), and H\textsc{ii} regions -- which comprise the most populated phases reported in \citet{cepedaarroita2025}.

Third, with a view to future AME analyses based on spinning dust theory, we ask how the SED can be represented efficiently within a general model framework. To this end, we develop a spectral fitting approach based on a moment expansion [see \cite{chluba2017rethinking}] with respect to the key parameters identified above, and we also propose a simple likelihood-free method for parameter inference.

The paper is organised as follows: Section~\ref{sec: SED var with env and grain} examines how spinning dust emission varies with environmental conditions and dust-grain properties, including a brief introduction to the models (\ref{sec: dust env model}), a case study to highlight general trends (\ref{sec: SED var case study}) and a global sensitivity analysis to identify the key parameters that dominate AME features (\ref{sec: gsa single config}). Building on this, Section~\ref{sec: ensemble analysis} explores the ensemble effects arising from a generic log-normal distribution over the key parameters, and compares the resulting Monte Carlo predictions with observational catalogues. In Section~\ref{sec: surrogate models}, we introduce surrogate modelling strategies for AME SED fitting and feature analysis. Section~\ref{sec: mom exp} presents the moment-expansion method, which removes the need for explicit distribution models, while Section~\ref{sec: moment emu} describes an emulation approach that directly maps observed AME features to model parameters. Finally, Section~\ref{sec: conclusion} summarises the work and presents the conclusions.
\section{SED Variations with Environment and Grain Morphology}
\label{sec: SED var with env and grain}

We now revisit the response of the spinning dust SED\footnote{By `spinning dust', we specifically mean the emission of a rotating electric dipole.} of single-type grains to variations in environmental conditions and grain morphological properties.

\subsection{Dust and ISM Environment Models}
\label{sec: dust env model}
Spinning electric dipole emission shows a more detailed and specific dependence on the interstellar medium (ISM) environment and the morphological model of the grain than the other radiative features of interstellar dust grains. 
Below, we summarise the models adopted for the analyses presented in this paper.

\paragraph*{Dust grain model.} 

We adopt the grain modelling approach of \citet{zhang2025spydust} (hereafter ZC25). 
The grain is modelled as a rigid triaxial rotor with principal moments of inertia $I_1$, $I_2$, $I_3$, reparameterized in terms of dimensionless shape parameters $\alpha$ and $\beta$ relative to a reference inertia $I_{\rm ref}$:
\begin{align*}
    \frac{1}{I_1}
    &=\frac{1+\alpha}{I_{\rm ref}}, &
    \frac{1}{I_2}
    &=\frac{1-\alpha}{I_{\rm ref}}, &
    \frac{1}{I_3}
    &=\frac{1+\beta}{I_{\rm ref}},
\end{align*}
where $\alpha$ describes deviations from circular symmetry within the grain plane, while $\beta$ characterizes the relative inertia of the axis perpendicular to the plane, and thus encodes the overall grain shape. In particular, $\beta>0$ corresponds to rod-like grains with a smaller axial moment of inertia, while $\beta<0$ corresponds to disc-like grains with a larger axial moment of inertia.

The parameter $\beta$ directly affects the rotational dynamics because the angular velocity components scale inversely with the corresponding moments of inertia. 
Less abstractly, the torque-free rotational dynamics of the dust grain is
\begin{align}
        \Dot{\theta} &= 
        \frac{L}{I_{\rm ref}}
        \alpha
        \sin{\theta}\sin{2\psi},
        \\
        \Dot{\phi} &= \frac{L}{I_{\rm ref}}
        \left(1-\alpha\cos{2\psi}
        \right),
        \\
        \Dot{\psi} &=
        \frac{L}{I_{\rm ref}}
        \cos{\theta}
        \left(\beta+\alpha\cos{2\psi}
        \right),
        \label{eq: angular frequecies definition}
\end{align}
where $\{\theta, \phi, \psi\}$ are the Euler angles describing the rotations of the grain body (see, for example, ZC25 for a detailed discussion).
As a result, for fixed angular momentum ($L$), rod-like grains ($\beta>0$) rotate faster about the symmetry axis, while disc-like grains ($\beta<0$) rotate more slowly about that axis. This modifies both the characteristic rotational frequencies and the distribution of rotational energy among modes, and therefore shapes the spectrum of rotational emission.

The above model is implemented in \textsc{SpyDust} as follows. Once the grain geometry is specified, assuming either a disklike or ellipsoidal grain with $\alpha=0$\footnote{As explained in ZC25, the labels of the principal moments of inertia can be permuted to minimise $\alpha$, thereby maximising the validity of this approximation.}, \textsc{SpyDust} computes the volume-equivalent radius of the grain. 
This radius serves as the effective grain size parameter a, consistent with the definition adopted in \textsc{SpDust2}.
This correspondence enables \textsc{SpyDust} to naturally inherit the \textsc{SpDust2} framework, in which the grain size $a$ is the central parameter governing the rotational and emission properties, while \textsc{SpyDust} introduces an additional shape parameter $\beta$, which modifies the relation between angular momentum and rotational frequency and also affects the calculation of the grain's cross-sectional area.

More specifically, the model introduces a characteristic size, $a_2=6$~\AA, as the delimitation: grains with $a\ge a_2$ are modelled as ellipsoids (with spherical grains being the limiting case), whereas smaller grains are treated as elliptical cylinders representing disk-like morphologies. 
The planar-spherical structural transition was originally discussed in Appendix~A of \cite{draine2001infrared}, while the value $a_2=6$~\AA corresponds approximately to the size of a large PAH containing $\sim \, 100$ carbon atoms \citep{AHD09}.
As described above, the shape of a dust grain is defined by two factors: in-plane ellipticity $\alpha$, which we fix to $0$ for axially symmetric grains, and axial oblateness $\beta$.
The adopted parameter ranges are $a \in (3.5,\,35)$~\AA\ and $\beta \in (-0.47,\,0.5)$,\footnote{In this work, we use the grain size limits built into \textsc{SpDust}, but recognise that the adopted limits should be revisited and discussed in future work.}
where the lower size limit is physically well motivated by sublimation constraints, which prevent smaller grains from surviving in typical interstellar environments \citep{guhathakurta1989temperature}.

We restrict the dust grain parameters to the size and shape, $a$ and $\beta$, and their distributions (discussed in Sect.~\ref{sec: distribution model}). We do not consider more complex grain properties, such as the charge distribution. 
Also note that, for all the simulations, we consider only tumbling grains, as distinguishing between tumbling and non-tumbling grains has a negligible impact on the results discussed in this paper.
The electric dipole moment is modelled following \cite{DL98b} and \cite{AHD09}, in which the total dipole moment is assumed to consist of two components: an intrinsic dipole moment arising from the addition of individual molecular bond dipoles, and a typically subdominant contribution due to charge displacement relative to the grain centre of mass [see \cite{DL98b} for detailed discussion].
The intrinsic dipole moment is modelled as a zero-mean multivariate Gaussian random variable. The root-mean-square magnitude is assumed to scale with the square root of the total number of atoms:
\begin{equation}
    \sigma_{\mu} = \sqrt{\mathrm{N}_{\rm atm} } \, \mu_{\rm atm}
\end{equation}
where $\mu_{\rm atm}$ denotes the characteristic dipole moment per atom and $\mathrm{N}_{\rm atm}$ is the number of atoms.
A fiducial value of $\mu_{\rm atm}\approx 0.38$~Debye is conventionally adopted \cite{AHD09}. 
In this work, we also consider a broader range $\mu_{\rm atm} \in [0.2, 0.6]$~Debye (in Sect.~\ref{sec: gsa single config}) to account for plausible variations in grain composition and internal structure.

\paragraph*{ISM environment model.} 
Spinning dust emission occurs in various phases of the interstellar medium.
The ISM environment is generally characterised by the following parameters:
(1) total hydrogen number density, $n_{\rm H}$ ($\text{cm}^{-3}$); (2) gas temperature, $T$ (K); (3) radiation field intensity relative to the average interstellar radiation field, $\chi$; (4) hydrogen ionization fraction, $x_{\rm H}\equiv n_{\rm H+}/n_{\rm H}$;
(5) ionized carbon fractional abundance, $x_{\rm C}\equiv n_{\rm C+}/n_{\rm H}$;
(6) molecular hydrogen fractional abundance, $y\equiv 2 n_{\rm H_2}/n_{\rm H}$.
The environmental parameters are adapted from Table~1 of \citet{DL98b} and \citet{draine2011physics}, with specific values provided in a table later in the analysis.

\paragraph*{AME source catalogues.} 
In this work, we consider three broadly defined ISM environments: (1) the molecular clouds; (2) the dark clouds; and (3) a range of H\textsc{ii} phases. 
These phases are of particular interest because the AME catalogues contain the largest number of sources that are each dominated by a single one of these environments. 
The AME feature catalogues, extracted from \cite{cepedaarroita2025}, are presented in Appendix~\ref{Append: catalogues}.
Here we note on the catalogue caveats:
(1) The MC and DC classifications are based on optical catalogues and are subject to selection biases. It is more likely that they trace the densest cores of larger, mixed-density dust structures rather than well-defined, homogeneous ISM phases. In practice, the separation between MC and DC is ambiguous, with no clear observational criteria distinguishing the two. 
(2) Sources classified as H\textsc{ii} regions trace dust structures associated with ionising sources rather than the ionised gas alone. The H\textsc{ii} catalogue includes regions in which the ionised component is dominant, but in all cases, strong dust clouds are present nearby. As \cite{cepedaarroita2025} noted, selection effects favour intrinsically bright regions in the sample.

\subsection{Spectral Feature Variations Across Grain Size, Shape, and Environment}
\label{sec: SED variation study}

\begin{figure}
    \centering
    \includegraphics[width=0.75\linewidth]{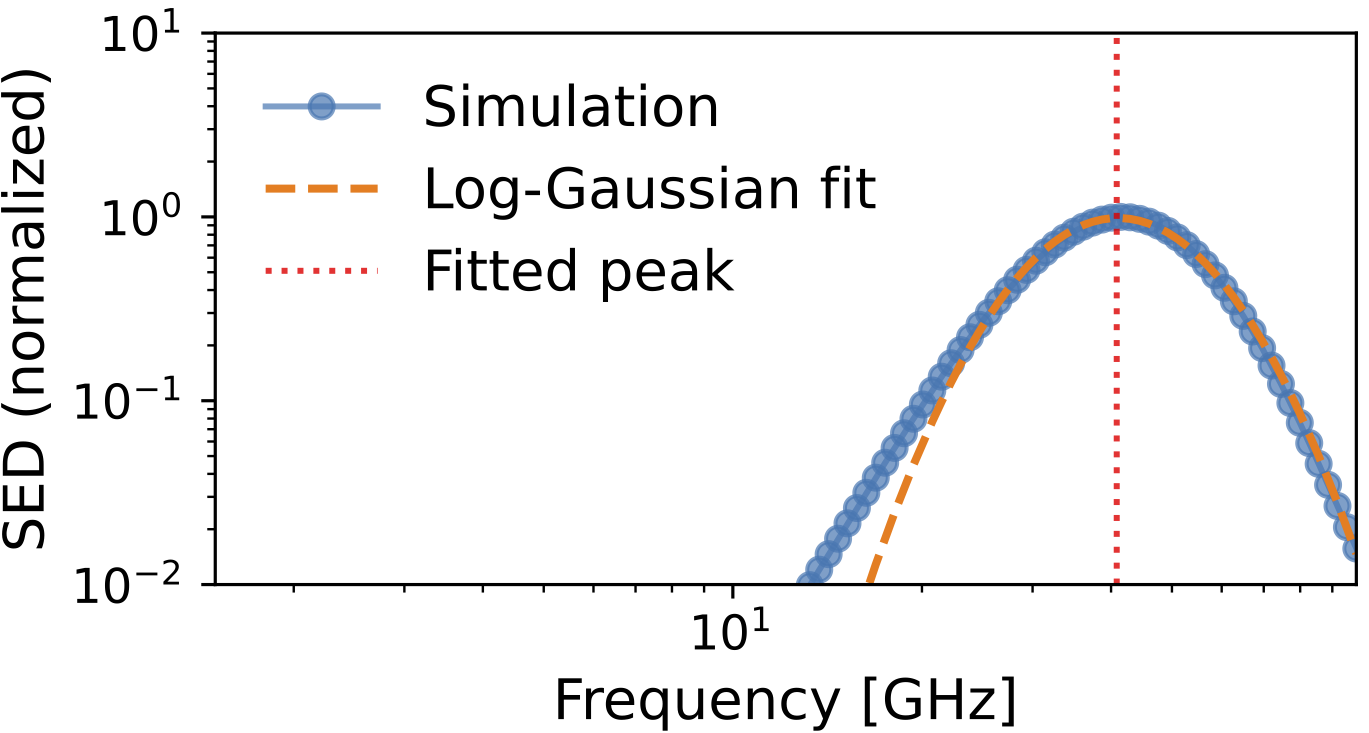}
    \caption{An illustrative example of fitting the SED with the log-Gaussian model (MC; $a=3.65\,$~\AA, $\beta=-0.224$).}
    \label{fig: log-Gaussian example}
\end{figure}

Given the above model assumptions, below we explore how the grain size and shape and ISM environmental properties shape the spectrum of the spinning dust emission. 
For the analysis, we simulate the SED using \textsc{SpyDust} and fit the SED to the log-Gaussian model \citep{Stevenson2014_lognormal, poidevin2023quijote} to extract the spectral features:
\begin{equation}
    I_\nu \propto \exp{\left[-\frac{1}{2}\left(\frac{\ln\nu-\ln{\nu_{\rm p}}}{W}\right)^2\right]},
    \label{eq: sed fit model}
\end{equation}
where $W$ characterises the SED width and $\nu_{\rm p}$ is the peak frequency.
We refer to this model as log-Gaussian because of its resemblance to a log-normal profile. However, it should be noted that this is not a strict log-normal distribution in the probabilistic sense, since $I_\nu$ is defined per unit frequency, rather than per logarithmic interval.
Fig.~\ref{fig: log-Gaussian example} provides an illustrative example of fitting the spinning dust emission to this model.

In this section, we examine two conditional cases to illustrate the general dependence of the spectral features. 
First, for fixed ISM environmental conditions, we study how the spectral features depend on the size and shape of the dust grains (Section~\ref{sec: varying grain}).
Second, for fixed dust grain parameters, we investigate how the spectral features vary with environmental conditions (Section~\ref{sec: varying env}). 
Next, in Sect.~\ref{sec: gsa single config}, we perform a global sensitivity analysis to systematically quantify the relative importance of each parameter in shaping the spectral signatures.

\begin{table}
\centering
\caption{ Idealised ISM phases adopted for the SED variation case study discussed in Section~\ref{sec: SED variation study}.
The specific values are adapted from Table~1 of \citet{DL98b} and \citet{draine2011physics}.
}
\begin{tabular}{lcccccc}
\hline
\hline
\textbf{Phase} &
\(\boldsymbol{n_{\rm H}}\) [cm\(^{-3}\)] &
\(\boldsymbol{T}\) [K] &
\(\boldsymbol{\chi}\) &
\(\boldsymbol{x_{\rm H}}\) &
\(\boldsymbol{x_{\rm C}}\) &
\(\boldsymbol{y}\) \\
\hline
Ideal MC & $300$ & $20$ & $0.01$ & $0$ & $10^{-4}$ & $0.99$ \\
Ideal DC & $10^{4}$ & $10$ & $10^{-4}$ & $0$ & $10^{-6}$ & $0.999$ \\
H\textsc{ii} (case 1) & $10$ & $8000$ & $10^3$ & $0.99$ & $10^{-5}$ & $0$ \\
H\textsc{ii} (case 2) & $10^4$ & $15000$ & $10^4$ & $0.999$ & $10^{-3}$ & $0$ \\
\hline
\end{tabular}
\label{table: ideal envs}
\end{table}

\subsubsection{Case Study of SED Response} 
\label{sec: SED var case study}

\begin{figure}
    \centering
    \includegraphics[width=\linewidth]{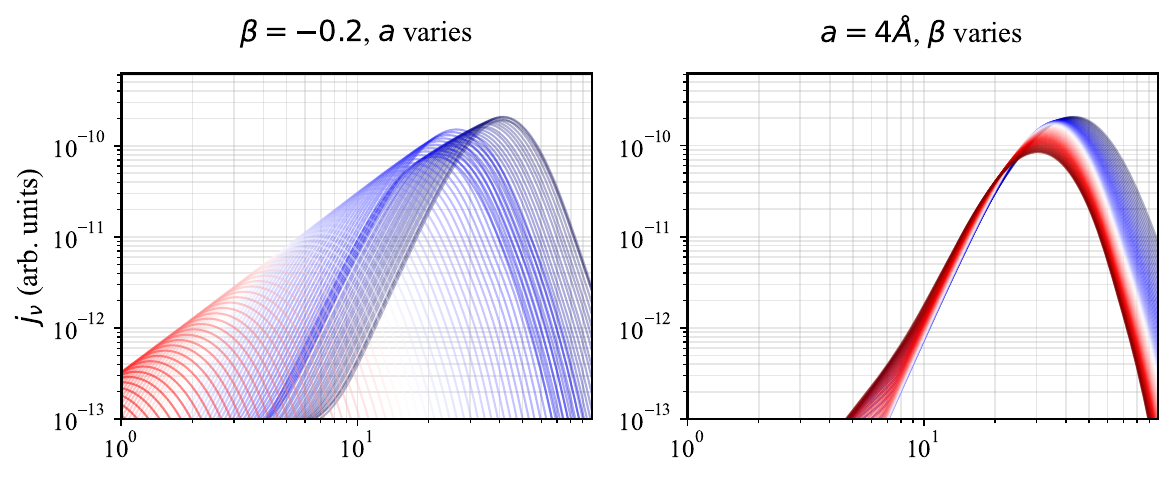}
    \caption{
    SED varies with grain parameters for a fixed MC environment. \emph{Left}: Given $\beta=-0.2$, the value of $a$ is sampled with a logarithmic scale from $3.5\text{ \AA}$ (blue) to $35\text{ \AA}$ (red). 
    The discontinuity arises from the discrete dust grain geometry model: small grains ($a < a_2$) are disk-like, while larger grains ($a \ge a_2$) are ellipsoidal.
    \emph{Right}: For a value of $a=4$~\AA, $\beta$ is uniformly sampled from $-0.47$ (blue) to $0.5$ (red).
    }
    \label{fig: SED varies with grain params}
\end{figure}

\begin{figure}
    \centering
    \includegraphics[width=\linewidth]{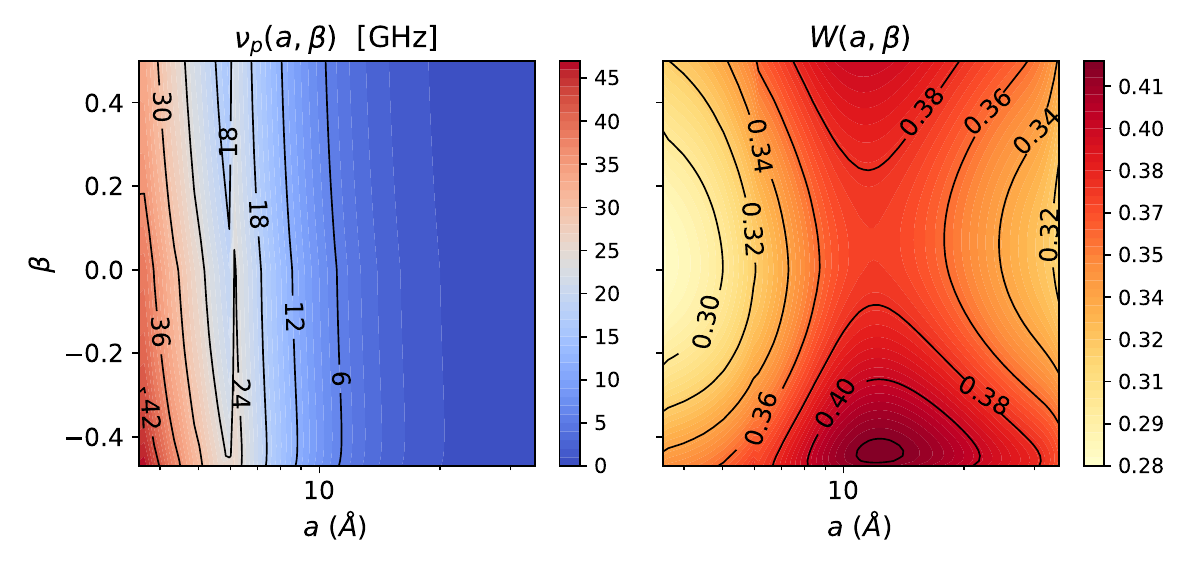}
    \caption{
    Heat maps of the peak frequency $\nu_{\rm p}$ and spectral width $W$ as functions of grain size $a$ and shape parameter $\beta$ in an idealised MC environment.
    As a reference, the observed mean $\nu_{\mathrm{AME}}$ is $21.975\,\mathrm{GHz}$ with a standard deviation of $4.248\,\mathrm{GHz}$, while the mean $W_{\mathrm{AME}}$ is $0.592$ with a standard deviation of $0.106$ (see Table~\ref{tab: MC source catalogue}).
    }
    \label{fig: heat map}
\end{figure}
\paragraph{Varying grain size and shape.}
\label{sec: varying grain}
For a fixed ISM phase -- here exemplified by the idealised MC (given in Table~\ref{table: ideal envs}) -- we examine how the spinning dust SED varies with the dust grain model. 
Figure~\ref{fig: SED varies with grain params} illustrates the changes in the SED when one grain parameter, $a$ (or $\beta$), is held fixed while the other, $\beta$ (or $a$), is varied. Variations in $a$ lead to a significant shift in the peak frequency, whereas changes in $\beta$ have little effect on the peak location but induce modest changes in the spectral width.
In general, smaller grain sizes correspond to higher peak frequencies.

These trends are more clearly seen in Fig.~\ref{fig: heat map}, which depicts the SED feature heat maps ($\nu_{\rm p}$ and $W$) as functions of $a$ and $\beta$. The predominantly horizontal structure of the heat maps indicates that $a$ governs the variation of the peak frequency $\nu_{\rm p}$, while both $a$ and $\beta$ contribute to the variation of the spectral width $W$.
The trend of $\nu_{\rm p}$ is also evident in Figure~1 of \cite{hensley2017modeling} and Figure~9 of \cite{AHD09}, both of which show a systematic shift of the peak frequency toward higher values for smaller grain sizes.

\paragraph{Varying ISM Environment.} 
\label{sec: varying env}

\begin{figure}
    \centering
    \includegraphics[width=\linewidth]{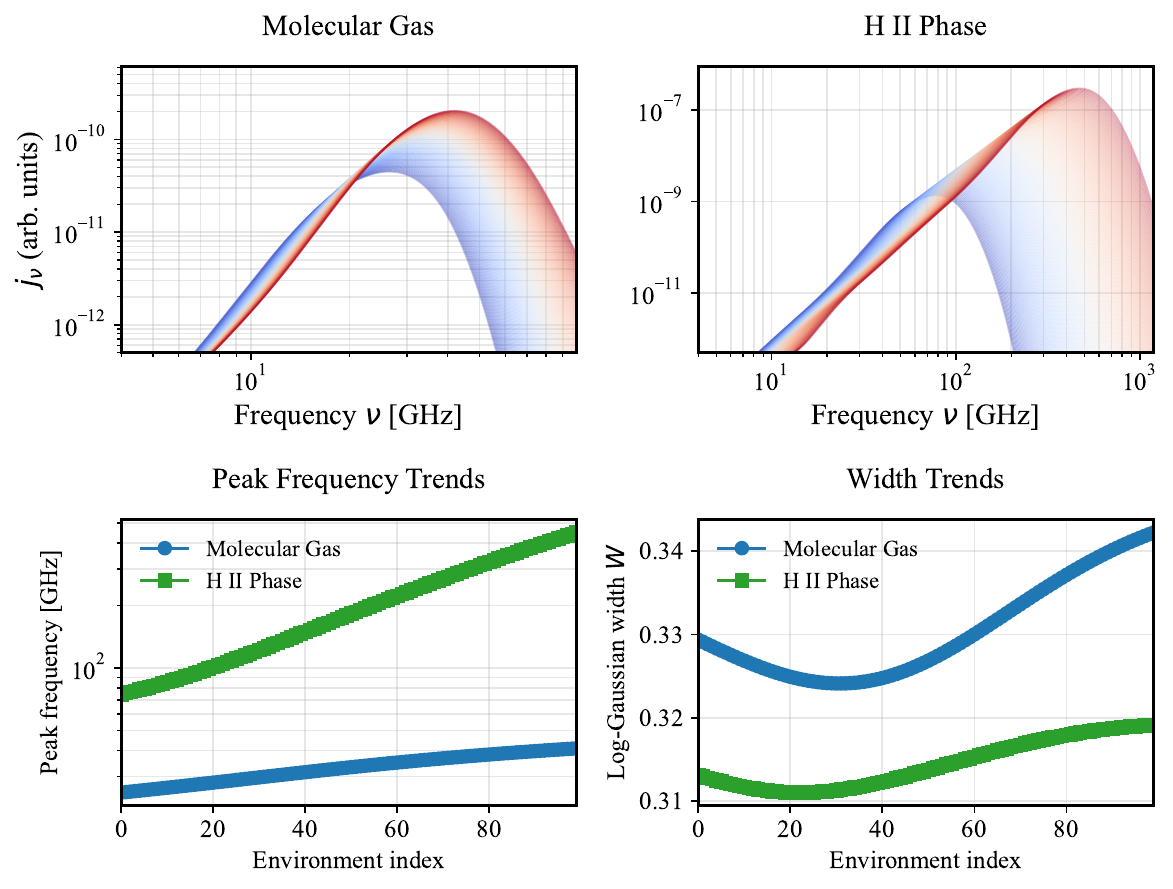}
    \caption{Spinning dust SEDs in varying ISM environment conditions for a fixed grain size of $a=4\text{ \AA}$, shape of $\beta=-0.427$ and dipole moment $\mu_{\rm atm}=0.38$ Debye.
    The coolwarm color scale represents the parameter variation. \emph{Top-left}: from the ideal dark cloud (blue) to the ideal molecular cloud (red); \emph{Top-right}: from H\textsc{ii} case~1 (low-density) to case~2 (high-density).
    The bottom panels show how the SED features change with environmental variation [as defined in eq~(\ref{eq: env interpolation})].}
    \label{fig: varying env}
\end{figure}

The emission of spinning dust exhibits complex, high-dimensional dependency on the ISM environment. In order to intuitively perceive the variance of SED features in response to a changing environment, we use a single parameter to interpolate between similar phases and model a wide range of environments.

Specifically, we interpolate all environmental parameters between the typical values of the idealised ISM phases (i.e. from ideal DC to ideal MC for molecular gas, and from low-density to high-density cases for H\textsc{ii}), as listed in Table~\ref{table: ideal envs}. All parameters are varied simultaneously using a single interpolation parameter on a logarithmic scale, except for the vanishing parameters.
As an example, for the temperature of molecular gas, the $i$-th sample satisfies
\begin{equation}
\ln T_i = \ln T_{\rm DC} + \frac{i}{N}(\ln T_{\rm MC} - \ln T_{\rm DC}),    
\label{eq: env interpolation}
\end{equation}
where $T_{\rm MC}=20$~K and $T_{\rm DC}=10$~K are idealised temperatures for each phase, and $N=100$ is the total number of samples. 
This interpolation is \emph{not} intended to represent a physical thermodynamic transition. It is primarily for parametric convenience rather than being physically motivated.
For example, the DC-MC transition involves non-linear processes such as self-shielding, chemical reactions and radiative transfer.

Figure~\ref{fig: varying env} shows the spinning dust SEDs of single-size-shape grains as the ISM environment varies. Environmental changes produce a wide range of peak frequencies, particularly for the considered range of H\textsc{ii}.
As the environment transitions from the ideal DC to MC, or from H\textsc{ii} case~1 to case~2, the peak frequency $\nu_{\rm p}$ increases noticeably. The spectral width $W$ also generally increases, but only moderately, remaining between $0.31$ and $0.34$.

\vspace{2em}
Based on the above case studies of grain models and ISM environments, we make the following general observations:
\begin{itemize}
    \item Peak frequency trend: The peak frequency is strongly affected by both the grain size $a$ and the environmental conditions. In contrast, variations in $\beta$ have a much smaller impact on the peak frequency.
    \item Width trend: Both the grain model (i.e., $a$ and $\beta$) and the environment contribute to changes in the spectral width. In particular, more disk-like or rod-like grains tend to produce a larger variation in width. 
    Overall, however, the width remains relatively constrained, typically between $0.3$ and $0.4$.
    \item Comparison with observations: If we compare the single-size, single-shape, single-environment spinning dust SEDs with the catalogue data presented in the Appendix~\ref{Append: catalogues}, we find that the observed spectral widths are systematically broader than those predicted by these idealised models, by a factor of $1.7 $ $(\approx 0.6/0.35)$. As for the peak frequency, the catalogue values for DC and MC lie within the model predicted range, whereas for H\textsc{ii} regions the observed peak frequencies are systematically lower than the model predictions.
\end{itemize}
The last point drives us to test the consistency between the model and observations in a more systematic way, beyond single configurations. 
With only two observational quantities and around ten model parameters, it is not possible to carry out a detailed validation, particularly when considering the arbitrary distribution effect over these parameters. 
Instead, we will first single out the few parameters that dominantly drive the SED variations (in Sect.~\ref{sec: gsa single config}) by performing a global sensitivity analysis. Then, by adopting a general distribution model for the key parameters (see Sect.~\ref{sec: ensemble analysis}), we can determine whether the observed catalogue lies within the range permitted by the model.

\subsubsection{Key Parameters Governing SED Variations}
\label{sec: gsa single config}

\begin{table*}
\centering
\caption{
Parameter ranges of the three broadly defined ISM phases considered in the global sensitivity analysis. Parameter ranges are estimated around the idealized phase values in Table~\ref{table: ideal envs}.
}
\begin{tabular}{lccccccccc}
\hline
\hline
\textbf{Phase} &
\(\boldsymbol{n_{\rm H}}\) [cm\(^{-3}\)] &
\(\boldsymbol{T}\) [K] &
\(\boldsymbol{\chi}\) &
\(\boldsymbol{x_{\rm H}}\) &
\(\boldsymbol{x_{\rm C}}\) &
\(\boldsymbol{y}\) &
\(\boldsymbol{a}\) [\AA] &
\(\boldsymbol{\beta}\) &
\(\boldsymbol{\mu_{\rm atm}}\) [Debye]
\\
\hline
MC & $10^{2}$ -- $10^{3}$ & $15$ -- $30$ & $10^{-3}$ -- $10^{-1}$ & $0$ & $10^{-5}$ -- $10^{-3}$ & $0.99$ -- $0.999$  & $3.5$ -- $35$  & $-0.47$ -- $0.5$ & $0.2$ -- $0.6$\\
DC & $10^{3}$ -- $10^{4}$ & $10$ -- $20$ & $10^{-5}$ -- $10^{-3}$ & $0$ & $10^{-7}$ -- $10^{-5}$ & $0.99$ -- $0.999$ & $3.5$ -- $35$  & $-0.47$ -- $0.5$ & $0.2$ -- $0.6$\\
H\textsc{ii} & $10$ -- $10^4$ & $8000$ -- $15000$ & $10^3$ -- $10^4$ & $0.99$ -- $0.999$ & $10^{-5}$ -- $10^{-3}$ & $0$ & $3.5$ -- $35$  & $-0.47$ -- $0.5$ & $0.2$ -- $0.6$\\
\hline
\end{tabular}
\label{table: environment}
\end{table*}

\begin{table*}
    \caption{Global sensitivity analysis of grain and environmental parameters for the two SED features, $\nu_{\rm p}$ and $W$, evaluated across all ISM phases. The definitions and methodological details of these metrics are provided in Appendix~\ref{Append: gsa concepts}.}
    \begin{tabular}{lccccccccccc}
    \toprule
    Feature & Parameter & MI & dCor & PermMean & PermStd & $S_1$ & $S_1^{\rm conf}$ & $S_T$ & $S_T^{\rm conf}$ & 1/ARD\_LS & AggRank \\
    \midrule
     \multirow{8}{*}{$\nu_{\rm p}$ (MC)} & $a$ & 0.405 & 0.698 & 1.450 & 0.061 & 0.749 & 0.075 & 0.836 & 0.077 & 0.424 & 1.200 \\
     & $x_{\rm C}$ & 0.102 & 0.330 & 0.649 & 0.029 & 0.058 & 0.027 & 0.100 & 0.019 & 0.836 & 2.000 \\
     & $\mu_{\rm atm}$ & 0.021 & 0.173 & 0.196 & 0.010 & 0.071 & 0.024 & 0.100 & 0.011 & 0.153 & 3.400 \\
     & $n_{\rm H}$ & 0.028 & 0.144 & 0.087 & 0.006 & 0.017 & 0.014 & 0.029 & 0.006 & 0.274 & 4.000 \\
     & $\beta$ & 0.006 & 0.094 & 0.085 & 0.005 & 0.041 & 0.018 & 0.048 & 0.006 & 0.149 & 5.200 \\
     & $\chi$ & 0.000 & 0.027 & 0.003 & 0.000 & 0.002 & 0.004 & 0.002 & 0.000 & 0.299 & 5.700 \\
     & $T$ & 0.000 & 0.022 & 0.001 & 0.000 & 0.001 & 0.004 & 0.002 & 0.000 & 0.015 & 7.100 \\
     & $y$ & 0.017 & 0.015 & 0.000 & 0.000 & -0.000 & 0.000 & 0.000 & 0.000 & 0.010 & 7.400 \\
     \midrule
     \multirow{8}{*}{$W$ (MC)} & $x_{\rm C}$ & 0.350 & 0.639 & 0.851 & 0.035 & 0.216 & 0.044 & 0.264 & 0.029 & 0.800 & 1.200 \\
     & $a$ & 0.260 & 0.588 & 0.661 & 0.024 & 0.560 & 0.063 & 0.656 & 0.054 & 0.537 & 1.800 \\
     & $\beta$ & 0.188 & 0.210 & 0.128 & 0.006 & 0.043 & 0.021 & 0.067 & 0.008 & 0.353 & 3.400 \\
     & $n_{\rm H}$ & 0.013 & 0.131 & 0.052 & 0.003 & 0.017 & 0.018 & 0.041 & 0.006 & 0.360 & 4.800 \\
     & $\mu_{\rm atm}$ & 0.022 & 0.065 & 0.009 & 0.001 & 0.003 & 0.018 & 0.051 & 0.010 & 0.235 & 5.200 \\
     & $T$ & 0.018 & 0.125 & 0.047 & 0.003 & 0.032 & 0.019 & 0.045 & 0.005 & 0.119 & 5.400 \\
     & $\chi$ & 0.000 & 0.041 & 0.004 & 0.000 & -0.003 & 0.008 & 0.010 & 0.002 & 0.381 & 6.300 \\
     & $y$ & 0.000 & 0.020 & 0.000 & 0.000 & 0.000 & 0.000 & 0.000 & 0.000 & 0.010 & 7.900 \\
     \midrule
     \multirow{8}{*}{$\nu_{\rm p}$ (DC)} & $a$ & 0.588 & 0.815 & 1.580 & 0.045 & 0.811 & 0.070 & 0.842 & 0.058 & 0.540 & 1.200 \\
     & $\beta$ & 0.023 & 0.158 & 0.209 & 0.008 & 0.085 & 0.026 & 0.090 & 0.010 & 0.690 & 2.200 \\
     & $x_{\rm C}$ & 0.021 & 0.168 & 0.237 & 0.009 & 0.065 & 0.021 & 0.061 & 0.007 & 0.196 & 2.800 \\
     & $n_{\rm H}$ & 0.016 & 0.118 & 0.045 & 0.004 & 0.008 & 0.010 & 0.016 & 0.004 & 0.345 & 4.000 \\
     & $\mu_{\rm atm}$ & 0.011 & 0.094 & 0.031 & 0.003 & 0.009 & 0.011 & 0.017 & 0.003 & 0.084 & 5.000 \\
     & $T$ & 0.000 & 0.060 & 0.041 & 0.002 & 0.009 & 0.011 & 0.012 & 0.002 & 0.059 & 6.100 \\
     & $\chi$ & 0.005 & 0.021 & -0.000 & 0.000 & -0.000 & 0.000 & 0.000 & 0.000 & 0.010 & 7.100 \\
     & $y$ & 0.000 & 0.019 & -0.000 & 0.000 & -0.000 & 0.000 & 0.000 & 0.000 & 0.010 & 7.600 \\
     \midrule
     \multirow{8}{*}{$W$ (DC)} & $\beta$ & 1.005 & 0.484 & 1.733 & 0.062 & 0.762 & 0.069 & 0.796 & 0.060 & 0.593 & 1.000 \\
     & $a$ & 0.071 & 0.203 & 0.094 & 0.005 & 0.125 & 0.030 & 0.145 & 0.020 & 0.519 & 2.000 \\
     & $n_{\rm H}$ & 0.053 & 0.191 & 0.074 & 0.002 & 0.036 & 0.018 & 0.034 & 0.004 & 0.459 & 3.400 \\
     & $\mu_{\rm atm}$ & 0.059 & 0.137 & 0.049 & 0.003 & 0.018 & 0.015 & 0.024 & 0.003 & 0.241 & 4.200 \\
     & $x_{\rm C}$ & 0.015 & 0.078 & 0.014 & 0.001 & 0.010 & 0.016 & 0.036 & 0.007 & 0.405 & 4.400 \\
     & $T$ & 0.001 & 0.047 & 0.004 & 0.000 & 0.019 & 0.011 & 0.018 & 0.003 & 0.101 & 6.000 \\
     & $y$ & 0.000 & 0.020 & 0.000 & 0.000 & -0.000 & 0.000 & 0.000 & 0.000 & 0.010 & 7.400 \\
     & $\chi$ & 0.000 & 0.019 & -0.000 & 0.000 & 0.000 & 0.000 & 0.000 & 0.000 & 0.010 & 7.600 \\
     \midrule
     \multirow{8}{*}{$\nu_{\rm p}$ (H\textsc{ii})} & $a$ & 0.449 & 0.581 & 0.000 & 0.000 & 0.885 & 0.080 & 0.964 & 0.069 & 0.862 & 1.800 \\
     & $n_{\rm H}$ & 0.009 & 0.052 & 0.002 & 0.009 & 0.020 & 0.026 & 0.088 & 0.021 & 4.121 & 2.200 \\
     & $\mu_{\rm atm}$ & 0.006 & 0.028 & 0.007 & 0.007 & -0.000 & 0.002 & 0.000 & 0.000 & 0.052 & 4.000 \\
     & $\chi$ & 0.000 & 0.033 & -0.009 & 0.007 & 0.014 & 0.013 & 0.028 & 0.005 & 0.399 & 4.800 \\
     & $\beta$ & 0.000 & 0.027 & -0.000 & 0.006 & 0.004 & 0.012 & 0.028 & 0.005 & 0.301 & 5.000 \\
     & $T$ & 0.017 & 0.023 & -0.005 & 0.007 & 0.001 & 0.002 & 0.001 & 0.000 & 0.022 & 5.200 \\
     & $x_{\rm C}$ & 0.000 & 0.018 & 0.004 & 0.005 & 0.000 & 0.000 & 0.000 & 0.000 & 0.010 & 6.300 \\
     & $x_{\rm H}$ & 0.003 & 0.015 & -0.003 & 0.007 & -0.000 & 0.000 & 0.000 & 0.000 & 0.010 & 6.700 \\
     \midrule
     \multirow{8}{*}{$W$ (H\textsc{ii})} & $\beta$ & 0.871 & 0.414 & 1.365 & 0.043 & 0.543 & 0.064 & 0.658 & 0.071 & 0.574 & 1.400 \\
     & $a$ & 0.142 & 0.305 & 0.000 & 0.000 & 0.228 & 0.064 & 0.434 & 0.056 & 0.929 & 2.800 \\
     & $n_{\rm H}$ & 0.042 & 0.098 & 0.047 & 0.007 & 0.005 & 0.038 & 0.139 & 0.020 & 6.041 & 3.000 \\
     & $\chi$ & 0.017 & 0.129 & 0.048 & 0.008 & 0.008 & 0.024 & 0.075 & 0.017 & 0.526 & 3.600 \\
     & $\mu_{\rm atm}$ & 0.040 & 0.102 & 0.018 & 0.005 & 0.009 & 0.019 & 0.044 & 0.009 & 0.235 & 4.400 \\
     & $x_{\rm C}$ & 0.000 & 0.026 & 0.002 & 0.002 & -0.000 & 0.000 & 0.000 & 0.000 & 0.010 & 6.600 \\
     & $T$ & 0.000 & 0.016 & -0.002 & 0.002 & -0.001 & 0.002 & 0.001 & 0.000 & 0.030 & 7.100 \\
     & $x_{\rm H}$ & 0.000 & 0.021 & -0.001 & 0.003 & 0.000 & 0.000 & 0.000 & 0.000 & 0.010 & 7.100 \\
    \bottomrule
    \label{tab: gsa table 1}
    \end{tabular}
\end{table*}

In order to systematically quantify the influence of sampled environmental and grain parameters on two key features of the model SEDs (the peak frequency, $\nu_{\mathrm{peak}}$, and the spectral width, $W$), we search the parameter space by generating 5,000 random samples, independently drawn for each parameter and ISM phase. 
Table~\ref{table: environment} summarises the parameter ranges for each ISM phase.

We evaluate the importance of each parameter using these Monte Carlo samples and the corresponding SED feature values.  
This enables us to identify the parameters predominantly responsible for variations in spinning dust emission features.

\paragraph*{Method}
Because the forward model lacks a closed-form analytic expression, we assess global sensitivity using a set of complementary metrics, each capturing a different aspect of parameter influence. Below we summarise the metrics used in this analysis\footnote{The Python package \textsc{MC-post}, we developed for this analysis, is available at \href{https://zh-zhang.com/mcpost/}{https://zh-zhang.com/mcpost/}.}; detailed definitions and methodological descriptions are provided in Appendix~\ref{Append: gsa concepts}. 

First, we compute mutual information (MI) and distance correlation (dCor) directly from the sample set in order to detect non-linear dependencies.
Next, we calculate permutation importance values using a Random Forest model trained on the data. The mean of the permutation importance measurements (PermMean) is used as the estimate of importance, while the standard deviation (PermStd) characterises the uncertainty in the estimate.

We also use a surrogate-assisted variance decomposition approach. We fit a Gaussian process (GP) regression with an automatic relevance determination (ARD) kernel to the data (with the inputs scaled to $[0, 1]$). The ARD length scales provide an additional measure of importance, with shorter length scales indicating greater sensitivity. We then evaluate the GP surrogate on Saltelli-Sobol samples to estimate first-order ($S_1$) and total-order ($S_T$) Sobol indices, with uncertainties quantified via approximate 95\% bootstrap confidence intervals ($S_1^{\mathrm{conf}}$ and $S_T^\mathrm{conf}$).
Finally, AggRank, the combination of various feature importance metrics as a unified ranking, provides an aggregated consensus ranking, combining all the individual importance metrics.

\paragraph*{Results}

Table~\ref{tab: gsa table 1} summarises the global sensitivity measurements for the two feature observations across different phases. 
The diagnostics are consistent across several independent importance measures, including MI, dCor, PermMean, $S_1$, $S_T$ and 1/ARD\_LS. We can therefore draw the following observations.

First, we observe that for each target feature, two parameters are sufficient to account for the majority of the variance. In this analysis, we focus on $S_1$, the first-order Sobol index, which provides a variance-based measure of sensitivity. By summing the $S_1$ values of selected parameters, we obtain a lower bound on their total contribution to the output variance. For example, for $\nu_{\mathrm{p}}$ (MC), $S_1(a) + S_1(x_{\mathrm{C}}) = 0.807$ indicates that $a$ and $x_{\mathrm{C}}$ together account for at least $80.7\%$ of the total variance. This is a lower bound because the sum does not include contributions from interactions between $a$ and $x_{\mathrm{C}}$.

Taking a closer look, it can be seen that the parameter $a$ emerges as a consistent driver of both the spectral width and the parameter $\nu_{\mathrm{p}}$. For $\nu_{\mathrm{p}}$, $a$ alone represents the majority of variance, exhibiting the highest MI and dCor values, and ranking first under the importance of permutation. 
Depending on the ISM phase, the variance of the spectral width is more evenly shared between $a$ and another parameter.

We identify the two dominant drivers for each target quantity:
\begin{itemize}
    \item \textbf{MC phases}: both $\nu_{\mathrm{p}}$ and $W$ are primarily driven by $a$ and $x_\mathrm{C}$;
    \item \textbf{DC phases}, both $\nu_{\mathrm{p}}$ and $W$ are primarily driven by $a$ and $\beta$;
    \item \textbf{H\textsc{ii} phases}, $\nu_{\mathrm{p}}$ is primarily driven by $a$ and $n_{\mathrm{H}}$, while $W$  is mainly controlled by $\beta$ and $a$.
\end{itemize}
For clarity in the subsequent discussion, we focus on three parameters for each ISM phase and adopt a unified notation for the dominant set,
\begin{equation*}
    \text{dominant parameters}\defeq \{a, \beta, p\}.
\end{equation*}
The grain size and shape parameters always present
and $p$ denotes the most influential environmental parameter: $p=x_{\rm C}$ for DC and MC phases, and $p= n_{\rm H}$ for H\textsc{ii} phases.
In Sect.~\ref{sec: ensemble analysis}, we will consider the ensemble effects arising from the distributions of these parameters, while fixing the remaining ones.

To conclude this analysis, we note that the global sensitivity results depend on the chosen parameter ranges. In other words, changing the range of a parameter may lead to a different assessment of its importance. Nevertheless, the ranges considered here are conservative and physically well motivated.

\section{Broadening effect from ensembles}
\label{sec: ensemble analysis}

Section~\ref{sec: gsa single config} provides a global sensitivity analysis for the three phases of interest. Based on that, for each phase, we singled out a three-parameter combination, $\{a, \beta, p \}$, to account for the dominating variance of the spectral features for each phase. Here, $p=x_{\rm C}$ for DC and MC, and $p= n_{\rm H}$ for H\textsc{ii} phases.

In this section, we study how the distribution function over these parameters affects the spinning dust emission features. 
As the remaining parameters contribute negligible variation to the SED, we fix them and disregard their ensemble effect. 
Section~\ref{sec: distribution model} introduces a general form of the distribution model.
We will then compare the observation catalogues with the SED features predicted by the distribution model in Sect.~\ref{sec: theory vs observation}.

\subsection{Distribution model of grain size, shape, environment}
\label{sec: distribution model}

\begin{table}
    \centering
    \caption{Summary of the physical ranges assumed for the grain and environmental parameters, and the associated parameter ranges defined for their distribution models.
    }
    \label{tab: dist_model}
    \begin{tabular}{lccc}
        \hline
        \textbf{Parameter} & \textbf{Symbol}     & \textbf{Range}  \\
        \hline
        Size & $a$ (\AA)    & $(3.5,\, 35)$ \\
        Shape & $\beta$ & $(-0.47,\,0.50)$   \\
        Environment (MC) & $x_{\rm C}$  &  $(10^{-5}, 10^{-3})$  \\
        Environment (DC) & $x_{\rm C}$  &  $(10^{-7}, 10^{-5})$  \\
        Environment (H\textsc{ii}) & $n_{\rm H}$  &  $(10, 10^{4})$  \\
        \hline
        Size distribution & $\gamma$ & $(-3, 3)$  \\
        Size distribution & $a_0$ (\AA) & $(0.1, 100)$  \\
        Size distribution & $\sigma$ & $(0.01, 20)$  \\
        Shape distribution & $\Tilde{\beta}_0$ & $(0.05, 2 )$ \\
        Shape distribution & $\delta$ & $(0.001, 20)$ \\
        Environment distribution & $\lambda$ & $(0.001, 20)$ \\
        Environment distribution (MC) & $p_0$ & $(10^{-5}, 10^{-3})$ \\
        Environment distribution (DC) & $p_0$ & $(10^{-7}, 10^{-5})$ \\
        Environment distribution (H\textsc{ii}) & $p_0$ & $(10, 10^{4})$ \\
        \hline
    \end{tabular}
\end{table}

We take a toy-model, multiplicative form for the distribution of the three dominating parameters
\begin{equation}
    f(a, \beta, p) = f(a) f(\beta) f(p).
    \label{eq: dist general form}
\end{equation}
For the size distribution, $f(a)$, we incorporate two fundamental models, guided by previous models and physical intuition. The first is a power-law form, which is expected to follow an efficient grain deconstruction process, such as the shattering effect that occurs in supernova shocks \citep{jones1996grain_shattering, hirashita2009shattering, jones2013evolution}. The second is a log-normal form, arguably resulting from the accelerated growth rate predicted in supersonic turbulence \citep{mattsson2020galactic}.
To unify these within a single parametric framework, we adopt the following expression:
\begin{equation}
    \label{eq: size distribution}
    f(a) \propto 
    \exp\!\left\{ (\gamma - 1 ) \ln a -  
    \tfrac{\left(\ln a - \ln a_0 + \frac{\sigma^2}{2}\right)^2}{2\sigma^2} \right\},
\end{equation}
for $a_{\rm min} \leq a \leq a_{\rm max}$.
This model reduces to a pure log-normal distribution for $\gamma=0$, approaches a pure power-law distribution as $1/\sigma \to 0$, and interpolates between the two as a generalised model otherwise. 
When $\gamma = 0$, the parameter $a_0$ represents the mean grain size (if the distribution is unbounded), and $\sigma$ describes the dispersion (width) of the size distribution.

As for the shape and environment distributions, $f(\beta)$ and $f(p)$, due to a lack of detailed knowledge or observational constraints, we also adopt a simple log-normal distribution:  
\begin{equation}
    \label{eq: shape distribution}
    f({ \Tilde{\beta}}) \propto \exp\!\left\{ -\ln \Tilde{\beta}  -   
    \tfrac{\left(\ln \Tilde{\beta} - \ln \Tilde{\beta}_0 + \frac{\delta^2}{2}\right)^2}{2\delta^2} \right\},
\end{equation}
and
\begin{equation}
    \label{eq: env distribution}
    f(p) \propto \exp\!\left\{ -\ln p  -   
    \tfrac{\left(\ln p - \ln p_0 + \frac{\lambda^2}{2}\right)^2}{2\lambda^2} \right\}.
\end{equation}
Here $\Tilde{\beta}_0$, $\delta$, $p_0$ and $\lambda$ are all free parameters. All distribution functions are normalised, e.g. $\int f (a) \diff{a} = 1$.

To summarise, the overall grain size and shape distribution is defined by Eqs.~(\ref{eq: size distribution}) and (\ref{eq: shape distribution}) and parameterised by seven key parameters: $\{ \gamma, \ln{a_0}, \sigma, \Tilde{\beta}_0, \delta, p_0, \lambda \}$.
Note that this model does not represent a definitive or ``ground-truth'' description; nevertheless, exploring it allows us to investigate how variations in grain morphology and environmental distributions may influence the SED features.

\subsection{Observed AME Features versus the Ensemble Model}
\label{sec: theory vs observation}

\begin{figure*}
    \centering
    \includegraphics[width=\linewidth]{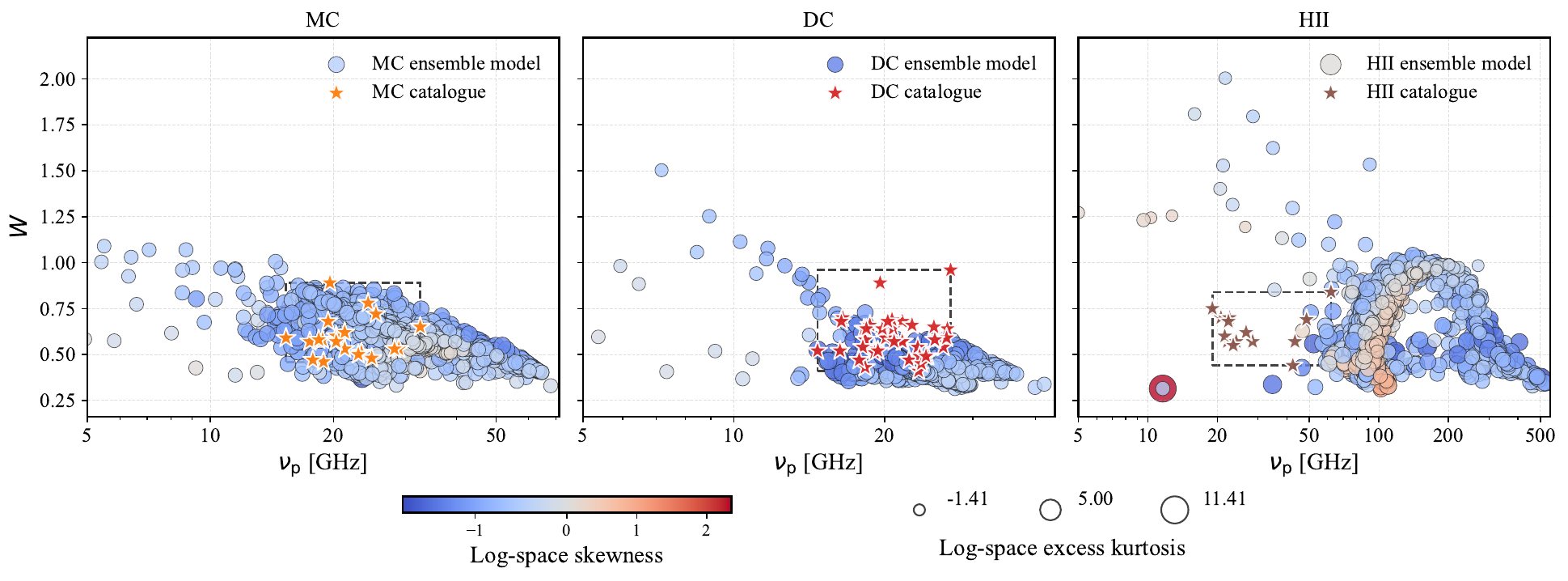}
    \caption{Spectral features of Monte Carlo realisations generated by sampling the grain-size, shape, and environmental distributions with the log-normal model and evaluating the corresponding spinning dust spectra for the three ISM phases (MC, DC, and H II). Each point represents one sample, with its position giving the peak frequency $\nu_{\rm p}$ and spectral width $W$. The colour encodes the log-space skewness $\gamma$, and the marker size reflects the log-space excess kurtosis $\kappa$. Star symbols show the observed AME catalogue for each phase. This comparison illustrates how each ISM phase occupies a distinct region of SED feature space defined by variations in the underlying distribution parameters. The dashed rectangular box indicates the parameter range covered by the observational catalogue, highlighting where the data lie within, or extend beyond, the domain predicted by the theoretical model.}
    \label{fig: obs vs full ensemble}
\end{figure*}

\begin{figure*}
    \centering
    \includegraphics[width=\linewidth]{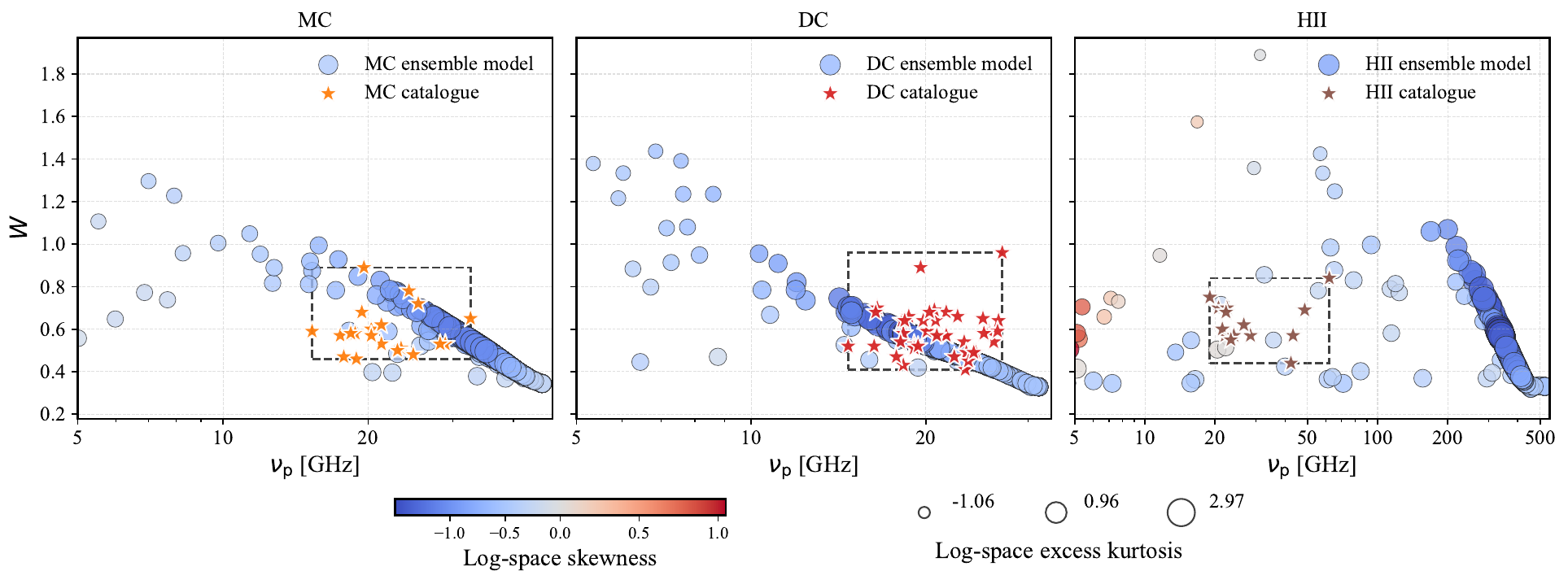}
    \caption{Reduced Monte Carlo sampling in which only the grain-size distribution parameters are varied, while $\beta$ and the environmental parameter $p$ are held fixed.
    Compared with Fig.~\ref{fig: obs vs full ensemble}, the modelled region in the $\nu_{\rm p}$–$W$ plane becomes substantially narrower: the allowed range of $W$ at fixed $\nu_{\rm p}$ (and vice versa) is greatly reduced. This shows that environmental variability, rather than the effects of grain size alone, is necessary to generate the full range of peak frequencies and spectral widths seen in the AME catalogue. The dashed rectangular box indicates the parameter range covered by the observational catalogue.}
    \label{fig: obs vs size ensemble}
\end{figure*}

Given the distribution model, i.e. Eqs.~(\ref{eq: dist general form}-\ref{eq: env distribution}), we set up the range for each parameter, as summarised in Table~\ref{tab: dist_model}. The parameter ranges are designed to cover distributions with centres ranging from near the lower to near the upper bounds, and with widths spanning from very broad to highly concentrated distributions.

We then generate samples of the distribution model by randomly sampling each distribution parameter independently.
The synthesised SED is computed as the Monte Carlo integral (the sum of distribution-weighted random points in the parameter space).
By computing the corresponding peak frequencies and spectral widths, we can determine the ranges of $\nu_{\mathrm{p}}$ and $W$ allowed by the distribution model within the assumed theoretical framework.

In addition to the peak frequency and spectral width, we compute the log-space skewness $\Tilde{\gamma}$ and \emph{excess} kurtosis $\kappa$ of the SED for each example. These are defined as follows:
\begin{equation}
\Tilde{\gamma} = \mu_3 / \mu_2^{3/2}, \quad \quad 
\kappa = \mu_4 / \mu_2^{2} - 3,
\end{equation}
where the moments are
\begin{equation}
    \mu_n =\sum_i p_i (x_i-\mu_1)^n, 
\end{equation}
with $x_i = \ln \nu_i$ and $p_i$ is the normalised weight given by
$p_i \propto I_\nu(x_i)\,\Delta x_i$, and $\mu_1$ is the log-space mean.
Both the skewness $\Tilde{\gamma}$ and the excess kurtosis $\kappa$ vanish when the SED is a perfect log-Gaussian.

Figure~\ref{fig: obs vs full ensemble} shows scatter plots of the random samples for each ISM phase. The position of each point encodes the peak frequency and spectral width, while the colour represents the skewness ($\Tilde{\gamma}$) of the log-space and the size of the marker reflects the excess kurtosis ($\kappa$) of the corresponding SED. The observed AME catalogue entries are overplotted as star markers in each panel.

Although higher-order summary statistics (including skewness and kurtosis) are not currently available from AME observations, they could be useful for more detailed validations of spinning dust models in the future. This is what we aim to demonstrate with the higher-order moments for the random samples shown in the figure.

Closer inspection of Fig.~\ref{fig: obs vs full ensemble} reveals different stories for each ISM phase.
\begin{itemize}
    \item For the MC phase, the observational catalogue lies fully within the region spanned by the assumed parameter distribution model combined with the spinning dust theory.

    \item For DC, most catalogue points are covered by the random samples, although some fall outside the allowed region. The sources of the two main outliers are identified in Table~\ref{tab: DC source catalogue}. 
    We note the substantial uncertainties in the width and peak frequency. Therefore, the discrepancy may be due to (1) a poor fit, for example, due to degeneracy with free-free emission, or (2) mild tension between the observations and either the theoretical model or the assumed parameter distributions.

    \item For H\textsc{ii}, we observe a significant discrepancy. The observational points are noticeably offset, suggesting that the observation catalogue genuinely falls outside the model's regime. 
    The discrepancy can be explained in two ways: First, observations of PAH from H\textsc{ii} regions -- see, for example, recent James Webb Space Telescope (JWST) studies of both extragalactic H\textsc{ii} regions \citep{sutter2024} and those in the Milky Way \citep{Peeters2024} -- reveal a significant suppression of PAH emission within H\textsc{ii} regions, indicating that the intense radiation fields associated with active star formation efficiently destroy small dust molecules. This suggests that small dust grains, which are responsible for spinning dust emission, are depleted in H\textsc{ii} environments.
    Second, in parallel to the deficit of H\textsc{ii} emissions, another known contributor to the discrepancy is the caveat associated with the H\textsc{ii} catalogue, which, as discussed in Sect.~\ref{sec: dust env model}, likely traces nearby dense dust clouds. 
    In other words, assuming that nanograins are significantly depleted in H\textsc{ii} regions, the observed H\textsc{ii} region may still exhibit AME originating from non-H\textsc{ii}  gas within the region. Consequently, model calculations based on ideal H\textsc{ii} conditions would be inappropriate, as the `H\textsc{ii}' observations are biased towards nearby non-H\textsc{ii} gas.  
    However, the observations do not invalidate the spinning dust mechanism itself, since AME associated with H\textsc{ii} regions likely arises predominantly from material surrounding the ionized gas, rather than from within the ionized region proper.
\end{itemize}

In Fig.~\ref{fig: obs vs size ensemble}, we present a reduced sampling experiment in which only the grain-size distribution parameters are varied, while the distributions of $\beta$ and $p$ (the environmental parameters) are held fixed. Compared with Fig.~\ref{fig: obs vs full ensemble}, the span of the sampling region is significantly narrower. In particular, for a given $\nu_{\mathrm{p}}$, the allowed range of $W$ is much smaller, and vice versa. This demonstrates that the ensemble effects arising from environmental variability\footnote{As we will show through the global sensitivity analysis, the role of $\beta$ is considerably less important.} are crucial for reproducing the breadth (i.e. the widely scattered pattern) seen in the observational catalogues.

We also perform a global sensitivity analysis using the random realisations of the distribution model, following the same procedure applied to the grain and environmental parameters in Sect.~\ref{sec: gsa single config}. The results are shown in Table~\ref{tab: gsa for ensemble model}.
In this case, however, no single parameter (or pair of parameters) emerges as universally dominant across all phases. The only consistent trend is that the $\beta$-distribution parameters, $\ln \tilde{\beta}_0$ and $\delta$, have negligible influence across all metrics and ISM phases.


\section{Surrogate Models for Spectral Fitting and Inference}
\label{sec: surrogate models}

In Section~\ref{sec: SED variation study}, we identified the dominant drivers of the SED features (peak frequency and width). In Section~\ref{sec: ensemble analysis}, we adopted a log-normal distribution model to explore how variations in these parameters shape the ensemble of synthesised SED features.

However, even a flexible log-normal model is still an explicit ansatz and may not fully capture the true structure of the parameter distributions. A key limitation is that the model is fully separable, with $a$, $\beta$ and $p$ treated as statistically independent. This removes any interaction between the parameters, and in the language of moment expansions (as detailed below), it eliminates spectral contributions from certain cross-derivative spectra.

To overcome this limitation, we consider an approach that does not assume an explicit distribution form. Instead, we work with a moment expansion [see examples in, e.g., \cite{chluba2017rethinking} and \cite{vacher2023high} for polarised radiation], relating the SED directly to the statistical moments of the underlying distributions. In Section~\ref{sec: mom exp}, we derive this relationship and show that the SED can be approximated using a small set of low-order moments provided that the true parameter distributions are reasonably concentrated.

Finally, in Section~\ref{sec: moment emu}, we introduce a practical technique based on \textsc{MomentEmu} \citep{zhang2025}\footnote{\url{https://github.com/zzhang0123/MomentEmu}} to connect the observed AME features to the distribution moments and hence infer these moments from data. This provides a pathway to inference that does not rely on prescribing a specific functional form for the parameter distributions.

\subsection{Moment expansion}
\label{sec: mom exp}

\begin{figure}
    \centering
    \includegraphics[width=\linewidth]{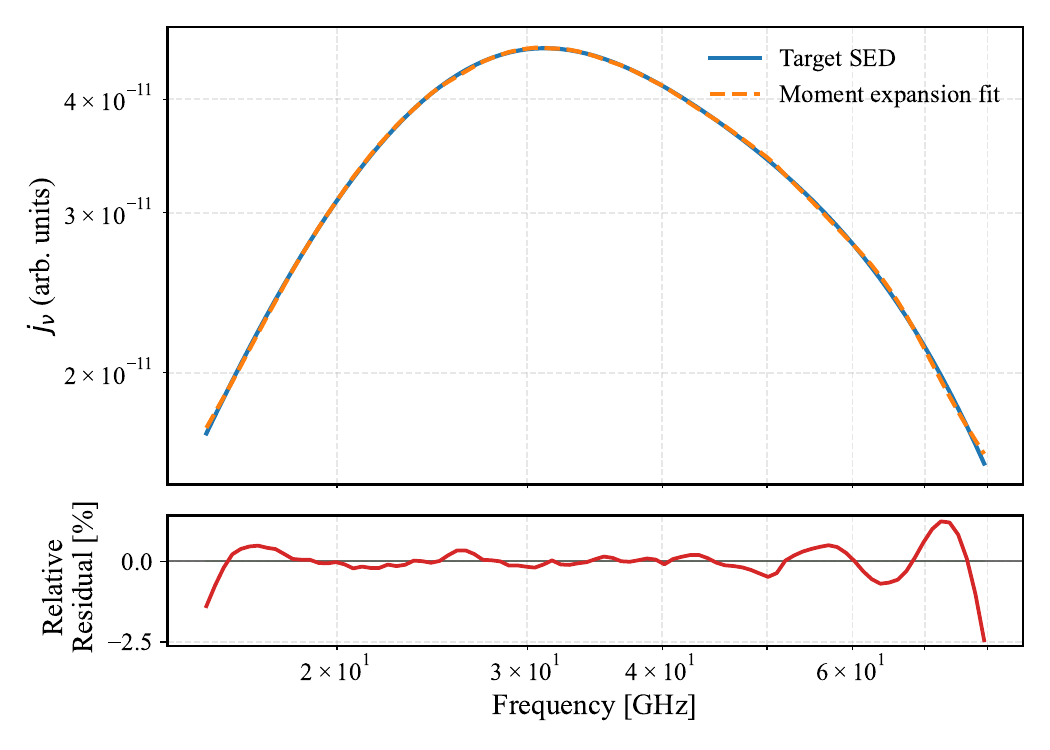}
    \caption{Example of a second-order moment expansion fit to a spinning dust SED generated from log-normal distributions in $a$, $\beta$, and $x_{\rm C}$ for the MC phase. \emph{Top}: raw and fitted SED. \emph{Bottom}: the fractional residual of the fit.}
    \label{fig: mom exp}
\end{figure}

The SED, synthesised over the distribution of the key parameters, can be written as follows:
\begin{equation}
    \Bar{I}_\nu
    =
    \iiint \diff{a} \diff{\beta} \diff{p} \;
    I_\nu (a, \beta, p) \, f  (a, \beta, p) 
    \equiv \langle I_\nu (a, \beta, p) \rangle.
\end{equation}
If we further perform a Taylor expansion of the SED with respect to these parameters, the synthesised SED can be expressed as
\begin{equation}
    \begin{split}
        \Bar{I}_{\nu} 
        = &
        I_{\nu} (\Bar{\boldsymbol{p}})
        + 
        \sum_i
        \left\langle
        p_i - \Bar{p}_i \right\rangle
        \partial_{ p_i} I_{\nu} \big\vert_{\Bar{\boldsymbol{p}}}
        \\
        + &
        \frac{1}{2!}\sum_i\sum_j
        \left\langle
        (p_i - \Bar{p}_i)(p_j - \Bar{p}_j) 
        \right\rangle
        \partial_{p_i} \partial_{p_j} I_{\nu}\big\vert_{\Bar{\boldsymbol{p}}}\\
        + &
        \frac{1}{3!}
        \sum_i\sum_j\sum_k
        \left\langle
        (p_i - \Bar{p}_i)(p_j - \Bar{p}_j)(p_k - \Bar{p}_k)  
        \right\rangle 
        {\partial_{p_i}\partial_{p_j}\partial_{p_k} I_{\nu}}\big\vert_{\Bar{\boldsymbol{p}}}\\
        +& \cdots
    \end{split}
    \label{eq: mom exp}
\end{equation}
where $\boldsymbol{p}$ denotes the full parameter vector, and $\Bar{\boldsymbol{p}}$ is the expansion pivot of the Taylor series.
It takes the form of a linear combination of the ``derivative spectra'', with the linear coefficients being the moments of the parameters centred at the pivot of Taylor expansion, $\Bar{\boldsymbol{p}}$.
Specifically, if we choose the pivot $\bar{\boldsymbol{p}}$ to be the mean of $\boldsymbol{p}$, the coefficients reduce to the conventional centred moments. In this case, the terms involving the first-order derivative spectra vanish.

In practice, we use Eq.~(\ref{eq: mom exp}) backwards: the explicit form of $f(a, \beta, p)$ is unknown, and the moment coefficients are treated as variables to be solved for, up to a chosen truncation order of the moment expansion, given the observed $\bar{I}_\nu$. 

Concretely, there are two operational approaches:
\begin{enumerate}
    \item Fixed-pivot method: We assume a fixed pivot, denoted by $\Bar{\boldsymbol{p}}$, and thus have a fixed set of derivative spectra that serve as linear basis functions. Then we linearly fit the coefficients of the derivative spectra.

    \item Pivot-fitting method: we fit both the pivot and the moment coefficients simultaneously. In this case, we impose the condition that the first-order derivative coefficients vanish, ensuring that the fitted pivot corresponds to the mean of $\boldsymbol{p}$.
\end{enumerate}

In our application, we do not have an analytical form for the derivative spectra. This means that we need to compute a numerical derivative, which can result in a long running time for a full Monte Carlo Markov Chain (MCMC) run using the pivot-fitting method (since the spectrum changes once the pivot is updated). Therefore, we opted for the fixed-pivot method. However, pivot-fitting may still be possible in the future if we can obtain fast surrogate models for the derivative spectra using, for example, \textsc{MomentEmu} \citep{zhang2025} or neural network methods.

We extend \textsc{SpyDust} by adding a moment-expansion module that fits a spinning dust SED to its moment expansion representation at a fixed pivot point with three parameters. Figure~\ref{fig: mom exp} illustrates an example: the SED generated from a random log-normal distribution in $a$, $\beta$, and $x_{\rm C}$ for the MC phase. When fitting the SED with a second-order moment expansion, the reconstruction closely matches the original spectrum, demonstrating that the moment expansion model provides an accurate low-order approximation.

\subsection{Emulation and likelihood-free inference with \textsc{MomentEmu}}
\label{sec: moment emu}

\begin{figure}
    \centering
    \includegraphics[width=0.95\linewidth]{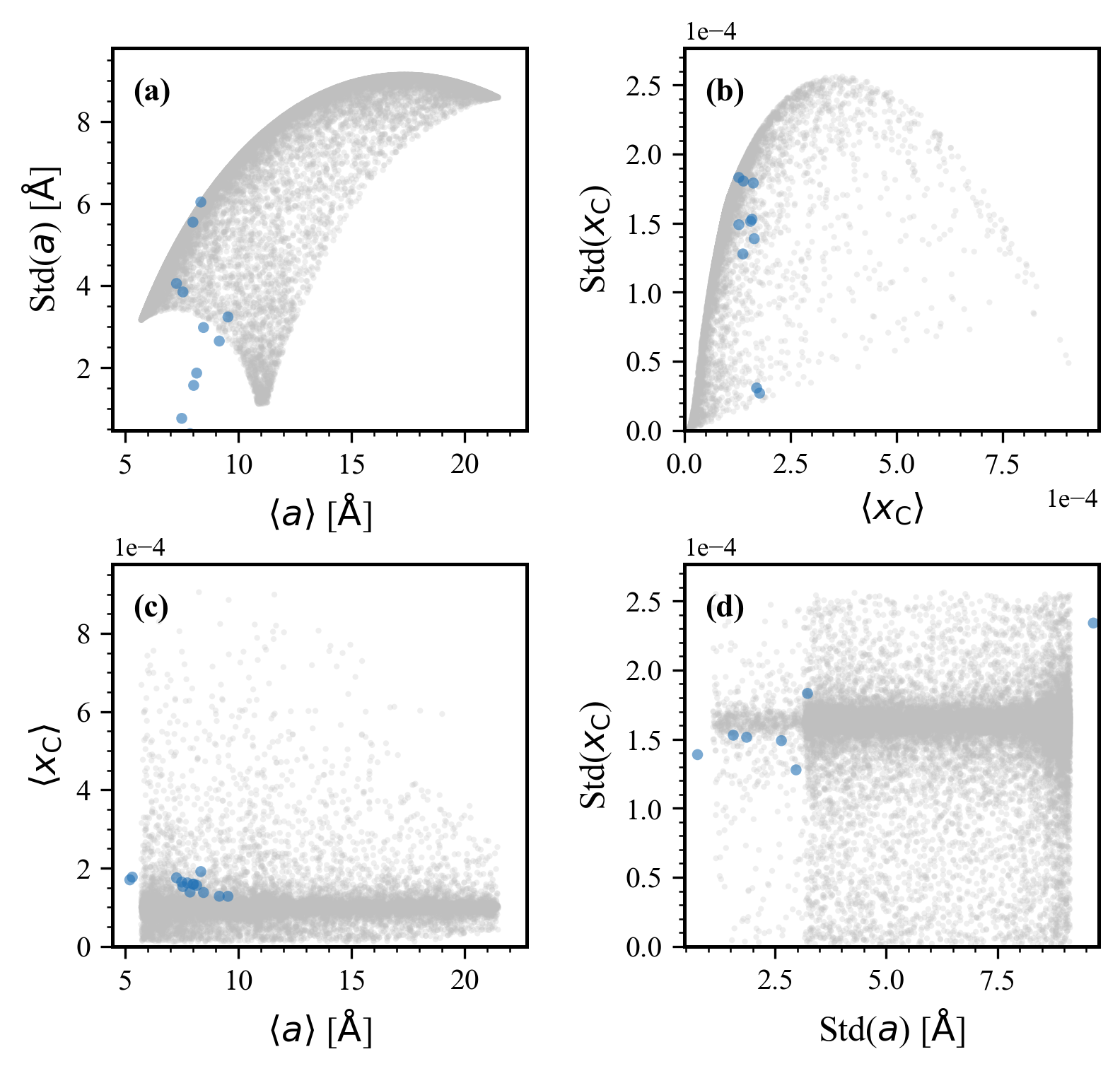}
    \caption{
    Inferred $\langle a\rangle$, $\mathrm{Std}(a)$, $\langle x_{\rm C}\rangle$, and $\mathrm{Std}(x_{\rm C})$ of the dust grain and environment distribution for the MC-phase AME catalogue using the \textsc{MomentEmu} polynomial surrogate.
    The grey background indicates the range covered by the training set.}
    \label{fig: mom inference}
\end{figure}

In the previous section, we formalised the moment expansion with respect to the key grain and environmental parameters, providing a linear representation of the SED in terms of moment coefficients and derivative spectra.

In this section, we propose an inference strategy that uses the observed AME features directly to constrain the parameters of the distribution model. The idea is to train an emulator on a Monte-Carlo-generated training set that links a sample of parameters to the corresponding AME features. The underlying distribution may be explicit, as in Sect.~\ref{sec: distribution model}, or implicit, such as a truncated set of moment, or more exactly, cumulant coefficients.

It is important to note that the amount of information that can be recovered about the distribution model is fundamentally limited by the number of observable AME features. For instance, it would be unrealistic to infer seven independent distribution parameters from just two observed features. Ideally, tighter constraints would necessitate the use of additional summary statistics, such as skewness or kurtosis. Another viable strategy is to reduce the dimensionality of the distribution model.

Here we adopt a simple proof-of-concept approach and use \textsc{MomentEmu} to construct a polynomial surrogate that maps observables to model parameters. 
As an illustrative example, we focus on the MC-phase AME catalogue. We select the four most influential parameters for the MC phase in the distribution model (see Table~\ref{tab: gsa for ensemble model}): $\lambda$ and $\ln{p_0}$ describe the distribution of $x_{\rm C}$, while $\gamma$ and $\sigma$ are associated with the distribution of $a$. 
We hold all other parameters fixed and allow the distribution functions to vary only along these four degrees of freedom.  For better physical interpretability, we label the distributions using their moments, $\langle a\rangle$, $\mathrm{Std}(a)$, $\langle x_{\rm C}\rangle$, and $\mathrm{Std}(x_{\rm C})$.\footnote{Std stands for the standard deviation, which is defined as the square root of the variance.} Our goal is to constrain these four physical quantities using the MC catalogue.

Despite adopting this reduced distribution model, the catalogue provides only two observable features, $\nu_{\mathrm{p}}$ and $W$. These two observables are insufficient to infer four independent distribution parameters. To address this limited information content, we impose additional constraints by fixing the skewness $\Tilde{\gamma}$ to $-0.2$ and assuming zero excess kurtosis  $\kappa$.\footnote{The assumed $\Tilde{\gamma}$ value lies well within the colour bar range shown in Fig.~\ref{fig: obs vs full ensemble}.}

We first generate a training set that establishes the correspondence between the distribution moments, $\{\langle a\rangle, \mathrm{Std}(a), \langle x_{\rm C}\rangle, \mathrm{Std}(x_{\rm C})\}$,  and the resulting SED features, $\{\nu_{\rm p}, W, \Tilde{\gamma}, \kappa\}$. Using these samples, \textsc{MomentEmu} learns a polynomial surrogate mapping from the SED features to the distribution moments:
\begin{equation}
    \{\nu_{\rm p}, W, \Tilde{\gamma}, \kappa\} 
    \;\longrightarrow\;
    \{\langle a\rangle, \mathrm{Std}(a), \langle x_{\rm C}\rangle, \mathrm{Std}(x_{\rm C})\}.
\end{equation}
Figure~\ref{fig: mom inference} shows the inferred distribution moments for the MC-phase catalogue obtained using the emulator. 
We note that some inferred values are invalid and therefore either do not appear in the plot or lie outside the training range. 
There are two likely explanations for this behaviour: (1) The mapping from SED features to distribution moments may be degenerate or numerically unstable. If so, additional observables would be needed to determine the distribution moments uniquely. (2) Uncertainties in the observations, or in the assumed skewness and kurtosis, may place some sources outside the regime sampled by the training set (i.e. outside the image space of the mapping).

We emphasise that this exercise is intended purely as a proof of concept and is limited by the assumed distribution model, fixed skewness and kurtosis, and observational uncertainties. In particular, the separable form adopted in Eq.~(\ref{eq: dist general form}) restricts the presence of certain spectral modes; in the language of moment expansions, some cross-moment modes vanish for this model. In future work, the use of more flexible distribution models (e.g. perturbative models with moments or cumulants) and improved AME feature measurements should enable this strategy to be applied more robustly.

\section{Conclusions and Discussions}
\label{sec: conclusion}

The apparent discrepancy between observed AME features and theoretical predictions \citep{cepedaarroita2025, fernandez2023quijote}, most notably the systematically lower spectral widths predicted by the theory, highlights the importance of critically re-examining the underlying assumptions and modelling strategies for spinning dust emission. Broadly, there are two avenues for theoretical interpretation: one involves refining the physical treatment of the spinning dust model itself, while the other focuses on revisiting the models of dust grains and astrophysical environments on which these calculations are based. In this paper, we assume the current spinning dust theory is valid and instead systematically study the impact of grain and environmental distributions on the synthesised SED features. Below, we summarise our key findings and conclusions.

\paragraph*{Key parameters.}
Through the global sensitivity analysis of the AME features (peak frequency and spectral width), we identify a three-parameter combination that dominantly drives the AME features in each ISM phase: $\{a, \beta, x_{\rm C}\}$ for MC and DC phases, and $\{a, \beta, n_{\rm H}\}$ for H\textsc{ii} regions. 
However, we reiterate that these results depend on the parameter ranges assumed for the Monte Carlo sampling. As our understanding of these ranges improves, the outcomes of the global sensitivity analysis may change accordingly.

\paragraph*{Ensemble broadening effects and comparisons with observations.}
Assuming log-normal distributions for the key parameters, we quantify the ensemble broadening effects on the synthesised SED features. Using Monte Carlo sampling, we compare the predicted AME features with observational catalogues. The tension between observations and theory is largely mitigated: 
(1) The MC-phase AME catalogue falls entirely within the model-predicted regime; (2) The DC catalogue exhibits mild discrepancies, although the two primary outliers can be explained by their large uncertainties; (3) The H\textsc{ii} catalogue shows significant discrepancies. In particular, the theoretical peak frequencies of H\textsc{ii} SEDs are systematically higher than the observed values. However, we reiterate that the H\textsc{ii} catalogue is also very likely to trace dense clouds nearby. 
This apparent discrepancy is consistent with the long-established deficit of PAH emission observed in H\textsc{ii} regions. Both point to a depletion of small dust grains, represented by PAHs, within the ionized gas.
Although it does not represent a tension between the general spinning dust scheme and observations, it suggests that revisiting H\textsc{ii} templates for spinning dust modelling is necessary.

In terms of the contributions of the individual parameters, the distributions of grain size and the dominant environmental parameter are the most significant. The variability of $\beta$ is less critical and can be fixed if necessary. The breadth of the observed features cannot be reproduced by variations in the grain size distribution alone, which highlights the importance of the variability of environmental distribution (see Fig.~\ref{fig: obs vs size ensemble}).

\paragraph*{Surrogate models and inference.}
We developed surrogate models of the SED and its features to enable efficient exploration and inference for fitting AME to spinning dust models. Two complementary approaches were employed:
\begin{enumerate}
    \item  Moment expansion: Expanding the SED in terms of derivatives with respect to key parameters provides a linearised representation that accurately reproduces the SED up to the moment-truncation order. This provides an inexpensive, computationally efficient way to fit AME spectra, study sensitivity, and predict SED variations in response to small changes in parameter distributions. The code is available as a \textsc{SpyDust} module.

    \item Polynomial emulator: We constructed a polynomial surrogate mapping from observable AME features ($\nu_{\rm peak}$, $W$, $\Tilde{\gamma}$, $\kappa$) to the four moments of the grain-size and environmental parameter distributions ($\langle a \rangle$, $\mathrm{Std}(a)$, $\langle x_{\rm C} \rangle$, $\mathrm{Std}(x_{\rm C})$). This enables fast, likelihood-free inference of distribution parameters directly from AME feature observations. Application to the MC-phase AME catalogue demonstrates that the emulator can recover meaningful constraints.
    Future work could strengthen this approach by adopting more flexible distribution models and improving AME feature measurements.
\end{enumerate}
Together, these surrogate approaches provide a flexible framework for linking observed AME spectra or features to the underlying properties of grains and the environment. They can also be extended to incorporate higher-order summary statistics or more complex distribution models in future studies.


\section*{Acknowledgements}
We would like to thank Dr Brandon Hansley for his insightful discussions on interpreting the nominal discrepancy in H\textsc{ii} regions. We are also grateful to colleagues in the Pan-Ex Journal Club and the RadioForegroundPlus update meeting for valuable feedback.
The results were obtained as part of a project that has received funding from
the RadioForegroundsPlus Project HORIZON-CL4-2023-SPACE-01, GA 101135036 (JARM, JC, RC, ZZ)
and the European Research Council (ERC) under the European Union's Horizon 2020 research and innovation programme (Grant agreement No. 948764; ZZ).  
JARM and RC acknowledge financial support from the Spanish MCIN/AEI/10.13039/501100011033, project ref. PID2023-151567NB-I00.
JC acknowledges support from the Fundaci\'{o}n Occident and the Instituto de Astrof\'{i}sica de Canarias under the Visiting Researcher Programme 2022-2025 agreed between both institutions.

\section*{Data Availability}

All code, data, and Jupyter notebooks necessary to reproduce the results presented in this paper are available in the associated GitHub repository: 
\url{https://github.com/SpyDust/SpyDust}.
The \textsc{MomentEmu} emulator referenced in Section~\ref{sec: moment emu} is available at:  \url{https://github.com/zzhang0123/MomentEmu}.
The \textsc{MC-post} package is available at:  \url{https://github.com/zzhang0123/mcpost}.



\bibliographystyle{mnras}
\bibliography{main} 



\appendix

\section{AME Spectral Feature Catalogues}
\label{Append: catalogues}
This appendix summarises the AME feature catalogues reported in \cite{cepedaarroita2025}: Table~\ref{tab: MC source catalogue} for the MC phase, Table~\ref{tab: DC source catalogue} for the DC phase, and Table~\ref{tab: Hii source catalogue} for the H\textsc{ii} phase.

\begin{table*}
\centering
\fontsize{8.5}{15.0}\selectfont

\caption{Catalogue of DC AME regions with best-fit parameters, identified by their central coordinates. 
The observed mean $\nu_{\mathrm{AME}}$ is $21.178\,\mathrm{GHz}$ with a standard deviation of $3.193\,\mathrm{GHz}$, while the mean $W_{\mathrm{AME}}$ is $0.590$ with a standard deviation of $0.108$.
A \textit{plus} sign indicates the presence of multiple additional environments of the same type, none of which are dominant.
$^\dagger$\,Regions that include a synchrotron component. 
$^\ast$\,Sources of the two main outliers in Fig~\ref{fig: obs vs full ensemble}.}
\label{tab: DC source catalogue} 
	
\begin{tabularx}{\linewidth}{@{\extracolsep{\fill}}l@{\hspace{1em}}c@{\hspace{1em}}c@{\hspace{1em}}c@{\hspace{1em}}p{6.5cm}@{}}
\toprule
Name & $A_{\mathrm{AME}}$ (Jy) & $\nu_{\mathrm{AME}}$ (GHz) & $W_{\mathrm{AME}}$ & \multicolumn{1}{l}{Main Sources (+Environment)} \\
\midrule

    G008.22$-$08.53 & $1.3 \pm 0.6$ & $25.7 \pm 3.8$ & $0.59 \pm 0.25$ & LDN 248 (DC) \\  
    G023.47+08.19 & $4.6 \pm 1.0$ & $18.3 \pm 3.2$ & $0.58 \pm 0.14$ & LDN462 (DC) \\ 
    G039.07$-$16.71 & $3.1 \pm 0.9$ & $16.5 \pm 3.2$ & $0.70 \pm 0.16$ & TGU H334 (DC)\,$\dagger$ \\ 
    G041.03$-$00.07 & $58.1 \pm 5.7$ & $21.8 \pm 0.7$ & $0.57 \pm 0.05$ & SDC G41.003-0.097 (DC)\,$\dagger$ \\ 
    G054.25+06.87 & $1.7 \pm 0.2$ & $18.2 \pm 1.1$ & $0.64 \pm 0.15$ & TGU H386 (DC) \\ 
    G057.66+00.09 & $24.6 \pm 2.1$ & $18.7 \pm 0.6$ & $0.66 \pm 0.06$ & DB 2003/05/06/07/09/12, TGU H398 (DC)\,$\dagger$ \\  
    G066.45$-$02.73 & $8.2 \pm 0.6$ & $21.7 \pm 0.9$ & $0.68 \pm 0.07$ & PGCC G066.53-02.87, DB 2140/41 (DC) \\  
    G078.44+00.17 & $42.0 \pm 10.4$ & $22.7 \pm 2.7$ & $0.66 \pm 0.18$ & DB 2403/04/09/15/23/27 (DC) \\ 
    G082.81$-$01.94 & $11.7 \pm 3.4$ & $23.3 \pm 3.1$ & $0.54 \pm 0.19$ & LDN 914, TGU H497, DB 2650/56/60/64 (DC) \\ 
    G094.76$-$01.52 & $6.9 \pm 1.2$ & $21.7 \pm 2.0$ & $0.57 \pm 0.13$ & LDN 1059 \,\textsuperscript{G094.47$-$01.53}\, (DC) \\ 
    G104.60+10.75 & $3.4 \pm 0.6$ & $26.2 \pm 1.9$ & $0.54 \pm 0.11$  & TGU H634 (DC) \\ 
    G106.44+12.39 & $2.5 \pm 0.2$ & $20.2 \pm 0.6$ & $0.59 \pm 0.06$ & LDN 1199, TGU H653 (DC) \\ 
    G107.29+19.10 & $1.5 \pm 0.3$ & $14.7 \pm 1.4$ & $0.52 \pm 0.11$ & TGU H667 (DC) \\ 
    G109.29+13.51 & $7.3 \pm 0.6$ & $16.4 \pm 0.6$ & $0.68 \pm 0.06$ & TGU H693, PGCC G109.11+13.26 (DC) \\ 
    G109.31+06.46 & $5.8 \pm 0.4$ & $24.4 \pm 0.7$ & $0.49 \pm 0.04$ & DB 3395/3397/3401/3389, LDN 1213/14 (DC) \\ 
    G110.72$-$00.48 & $9.0 \pm 1.9$ & $26.7 \pm 2.3$ & $0.64 \pm 0.15$ & DB 3435/33, TGU H700 (DC) \\ 
    G122.67+09.73 & $2.9 \pm 0.6$ & $16.3 \pm 1.4$ & $0.52 \pm 0.18$ & IREC 169, TGU H814, DB 3740 (DC) \\ 
    G124.29+02.57 & $11.8 \pm 1.0$ & $18.4 \pm 0.7$ & $0.64 \pm 0.09$ & LDN 1307, DB 3760/61/63, TGU H823 (DC) \\ 
    G124.91$-$03.81 & $3.4 \pm 0.4$ & $20.8 \pm 1.4$ & $0.64 \pm 0.14$ & LDN 1310 (DC) \\ 
    G127.67+13.88 & $1.3 \pm 0.2$ & $23.4 \pm 1.4$ & $0.41 \pm 0.07$ & TGU H853, DB 3797/98/99/3800/01 (DC) \\ 
    G129.15$-$00.07 & $2.3 \pm 0.7$ & $19.6 \pm 3.9$ & $0.89 \pm 0.22$ & LDN 1332/34/37, DB 3809/12 (DC)\,$\dagger$ $\ast$ \\ 
    G129.17$-$04.89 & $1.3 \pm 0.3$ & $18.3 \pm 1.7$ & $0.43 \pm 0.15$ & PGCC G129.19-04.84, TGU H858 (DC)\,$\dagger$ \\ 
    G150.16+09.36 & $3.5 \pm 0.6$ & $19.1 \pm 1.2$ & $0.53 \pm 0.12$ & PGCC G150.43+09.39, DB 4066 (DC)\,$\dagger$ \\  
    G155.87+05.08 & $2.2 \pm 0.3$ & $20.1 \pm 1.0$ & $0.59 \pm 0.09$ & LDN 1436/38, DB 4110/12/13/14/15 (DC)\,$\dagger$ \\ 
    G159.02$-$33.88 & $4.0 \pm 0.3$ & $19.8 \pm 1.1$ & $0.64 \pm 0.08$ & LDN 1454/53/58, DB 4162 (DC)\,$\dagger$ \\ 
    G164.48$-$05.63 & $1.6 \pm 0.2$ & $22.4 \pm 0.9$ & $0.47 \pm 0.05$ & DB 4245/51/52/55, LDN 1481, TGU H1116 (DC) \\ 
    G169.84$-$08.99 & $2.5 \pm 0.4$ & $20.9 \pm 1.4$ & $0.57 \pm 0.11$ & LDN 1496, TGU H1157, DB 4313 (DC) \\ 
    G171.79$-$00.09 & $4.2 \pm 0.3$ & $26.6 \pm 1.4$ & $0.59 \pm 0.08$ & TGU H1175/72/81 (DC) \\ 
    G174.27$-$13.81 & $3.5 \pm 0.3$ & $19.1 \pm 0.8$ & $0.51 \pm 0.07$ & LDN 1527 (DC)\,$\dagger$ \\ 
    G175.60$-$12.51 & $4.3 \pm 0.3$ & $18.1 \pm 0.6$ & $0.54 \pm 0.05$ & DB 4466, TGU H1211 (DC) \\ 
    G183.81$-$20.35 & $2.8 \pm 0.6$ & $27.1 \pm 8.9$ & $0.96 \pm 0.29$ & TGU H1315 (DC) $\ast$\\ 
    G192.41$-$11.51 & $12.9 \pm 0.7$ & $25.2 \pm 0.6$ & $0.58 \pm 0.04$ & $\lambda$ Orionis B30 \,\textsuperscript{G192.34$-$11.37}\, (DC) \\ 
    G194.77$-$15.71 & $12.4 \pm 0.5$ & $23.5 \pm 0.3$ & $0.49 \pm 0.02$ & $\lambda$ Orionis B223 (DC) \\ 
    G195.90$-$02.60 & $5.5 \pm 0.3$ & $23.7 \pm 0.6$ & $0.45 \pm 0.03$ & LDN 1591/92/93 (DC) \\ 
    G199.55$-$11.81 & $4.8 \pm 0.5$ & $25.1 \pm 3.8$ & $0.65 \pm 0.16$ & DB 4695, LDN 1602/03 (DC) \\  
    G216.31+09.85 & $1.2 \pm 0.3$ & $19.4 \pm 3.9$ & $0.52 \pm 0.15$ & TGU H1512, PGCC G216.04+09.86 (DC)\,$\dagger$ \\ 
    G225.91$-$00.44 & $5.2 \pm 0.5$ & $20.7 \pm 2.3$ & $0.69 \pm 0.11$ & DB 5087/90/91/92... (DC) \\ 
    G231.83$-$02.00 & $4.0 \pm 0.3$ & $17.8 \pm 0.7$ & $0.47 \pm 0.06$ & [K60] 201, TGU H1593, DB 5103 (DC) \\ 
    G247.60$-$12.40 & $3.1 \pm 0.2$ & $20.3 \pm 1.5$ & $0.68 \pm 0.07$ & TGU H1630, DB 5147/48 (DC) \\     
    G353.97+15.79 & $16.4 \pm 2.4$ & $24.2 \pm 1.2$ & $0.49 \pm 0.09$ & $\rho$ Ophiuchi East (DC) \\ 
\bottomrule
\end{tabularx}
\end{table*}

\begin{table*}
\centering
\fontsize{8.5}{15.0}\selectfont

    \caption{Catalogue of MC AME regions with best-fit parameters, identified by their central coordinates. 
    The observed mean $\nu_{\mathrm{AME}}$ is $21.975\,\mathrm{GHz}$ with a standard deviation of $4.248\,\mathrm{GHz}$, while the mean $W_{\mathrm{AME}}$ is $0.592$ with a standard deviation of $0.106$.
    A \textit{plus} sign indicates the presence of multiple additional environments of the same type, none of which are dominant. $^\dagger$\,Regions that include a synchrotron component.}
	\label{tab: MC source catalogue} 

    \begin{tabularx}{\linewidth}{@{\extracolsep{\fill}}l@{\hspace{1em}}c@{\hspace{1em}}c@{\hspace{1em}}c@{\hspace{1em}}p{6.5cm}@{}}
	\toprule

    Name & $A_{\mathrm{AME}}$ (Jy) & $\nu_{\mathrm{AME}}$ (GHz) & $W_{\mathrm{AME}}$ & \multicolumn{1}{l}{Main Sources (+Environment)} \\
	\midrule

    G010.84$-$02.59 & $12.9 \pm 2.9$ & $21.3 \pm 1.5$ & $0.53 \pm 0.15$ & GGD 27 (MC)\,$\dagger$ \\  
    G018.02+12.45 & $5.2 \pm 0.7$ & $24.3 \pm 2.5$ & $0.78 \pm 0.14$ & TGU H189, DB 0768 (MC)\,$\dagger$ \\ 
    G018.53+18.05 & $2.2 \pm 0.5$ & $18.9 \pm 2.9$ & $0.46 \pm 0.10$ & TGU H198, PGCC G018.23+18.06 (MC) \\  
    G040.10$-$35.50 & $3.2 \pm 1.1$ & $15.3 \pm 3.4$ & $0.59 \pm 0.19$ & MBM 47 (MC) \\
    G089.33+11.17 & $4.0 \pm 0.7$ & $17.5 \pm 0.8$ & $0.57 \pm 0.09$ & PGCC G089.54+11.28 (MC)\,$\dagger$ \\ 
    G096.89$-$29.75 & $1.1 \pm 0.2$ & $17.8 \pm 1.2$ & $0.47 \pm 0.11$ & PGCC G096.76-29.39 (MC)\,$\dagger$ \\ 
    G099.60+03.70 & $11.0 \pm 2.5$ & $32.6 \pm 2.9$ & $0.65 \pm 0.18$ & LDN1111 (MC) \\ 
    G108.89$-$52.21 & $0.8 \pm 0.2$ & $23.4 \pm 3.8$ & $0.51 \pm 0.25$ & PGCC G108.79-51.98 (MC) \\ 
    G124.39+30.21 & $2.2 \pm 0.4$ & $20.3 \pm 2.7$ & $0.60 \pm 0.11$ & Polaris Flare (MC)\,$\dagger$ \\ 
    G126.80$-$70.10 & $1.8 \pm 0.5$ & $25.4 \pm 5.1$ & $0.72 \pm 0.18$ & PGCC G127.36-70.07 (MC)\,$\dagger$ \\ 
    G133.27+09.05 & $4.7 \pm 0.7$ & $18.8 \pm 0.9$ & $0.58 \pm 0.08$ & LDN 1358/1355/1357 (MC)\,$\dagger$ \\ 
    G142.60+38.46 & $1.0 \pm 0.5$ & $19.4 \pm 4.0$ & $0.68 \pm 0.20$ & Ursa Major Complex (MC)\,$\dagger$ \\ 
    G154.03$-$39.80 & $2.1 \pm 0.4$ & $19.6 \pm 3.9$ & $0.89 \pm 0.18$ & PGCC G154.64-39.67 (MC)\,$\dagger$ \\ 
    G160.26$-$18.62 & $19.8 \pm 0.6$ & $24.8 \pm 0.5$ & $0.48 \pm 0.02$ & Perseus Molecular Cloud (MC) \\ 
    G170.60$-$37.30 & $21.8 \pm 0.6$ & $21.3 \pm 0.4$ & $0.62 \pm 0.03$ & MBM 16 (MC) \\ 
    G206.60$-$26.40 & $1.4 \pm 0.2$ & $23.0 \pm 4.0$ & $0.50 \pm 0.14$ & PGCC G206.55-26.17, DB 4820 (MC) \\ 
    G235.62+38.27 & $1.4 \pm 0.4$ & $18.4 \pm 3.5$ & $0.58 \pm 0.15$ & PGCC G235.60+38.28 (MC)\,$\dagger$ \\  
    G342.51+08.81 & $3.8 \pm 0.8$ & $20.3 \pm 4.5$ & $0.57 \pm 0.17$ & Lupus 5 (MC) \\ 
    G344.75+23.97 & $1.6 \pm 0.2$ & $28.9 \pm 1.8$ & $0.53 \pm 0.07$ & MBM 121/22 (MC) \\  
    G353.05+16.90 & $22.6 \pm 0.8$ & $28.2 \pm 0.8$ & $0.53 \pm 0.02$ & $\rho$ Ophiuchi (MC) \\ 

    \bottomrule
    \end{tabularx}
\end{table*}

\begin{table*}
\centering
\fontsize{8.5}{15.0}\selectfont

\caption{Catalog of H\textsc{ii} AME regions with best-fit parameters, identified by their central coordinates. 
The observed mean $\nu_{\mathrm{AME}}$ is $30.493\,\mathrm{GHz}$ with a standard deviation of $12.670\,\mathrm{GHz}$, while the mean $W_{\mathrm{AME}}$ is $0.632$ with a standard deviation of $0.097$.
A \textit{plus} sign indicates the presence of multiple additional environments of the same type, none of which are dominant.   $^\dagger$\,Regions that include a synchrotron component.}
\label{tab: Hii source catalogue} 

\begin{tabularx}{\linewidth}{@{\extracolsep{\fill}}l@{\hspace{1em}}c@{\hspace{1em}}c@{\hspace{1em}}c@{\hspace{1em}}p{6.5cm}@{}}
\toprule
Name & $A_{\mathrm{AME}}$ (Jy) & $\nu_{\mathrm{AME}} (GHz)$ & $W_{\mathrm{AME}}$ & \multicolumn{1}{l}{Main Sources (+Environment)} \\
\midrule

G028.79+03.49 & $8.2 \pm 1.9$ & $43.3 \pm 5.3$ & $0.57 \pm 0.16$ & W40 (H\textsc{ii}) \\ 
G037.79$-$00.11 & $62.1 \pm 7.8$ & $22.1 \pm 2.6$ & $0.57 \pm 0.11$ & W47 (H\textsc{ii}) \\  
G062.98+00.05 & $14.6 \pm 1.0$ & $21.5 \pm 0.8$ & $0.60 \pm 0.06$ & S89 (H\textsc{ii}) \\ 
G068.16+01.02 & $2.5 \pm 0.6$ & $28.4 \pm 4.5$ & $0.57 \pm 0.16$ & S98 (H\textsc{ii}) \\ 
G070.14+01.61 & $14.5 \pm 1.5$ & $26.6 \pm 1.3$ & $0.62 \pm 0.09$ & NGC 6857, S100 (H\textsc{ii}) \\  
G093.02+02.76 & $20.3 \pm 4.0$ & $20.7 \pm 1.9$ & $0.70 \pm 0.15$ & GAL093.06+2.81 (H\textsc{ii}) \\ 
G107.20+05.20 & $18.9 \pm 1.2$ & $24.1 \pm 0.9$ & $0.57 \pm 0.05$ & S140 (H\textsc{ii}) \\ 
G160.27$-$12.36 & $9.6 \pm 2.1$ & $62.0 \pm 13.2$ & $0.84 \pm 0.20$ & California Nebula \,\textsuperscript{G160.60$-$12.05}\, (H\textsc{ii}) \\ 
G173.56$-$01.76 & $5.8 \pm 0.7$ & $42.4 \pm 1.8$ & $0.44 \pm 0.05$ & IC 410 (H\textsc{ii}) \\ 
G173.62+02.79 & $12.0 \pm 0.5$ & $23.4 \pm 0.6$ & $0.55 \pm 0.03$ & S235 (H\textsc{ii}) \\ 
G176.90$-$00.41 & $10.8 \pm 0.9$ & $19.0 \pm 0.9$ & $0.75 \pm 0.08$ & IRAS 05331+3115 (H\textsc{ii}) \\ 
G192.60$-$00.06 & $7.3 \pm 0.4$ & $22.5 \pm 0.8$ & $0.70 \pm 0.07$ & S255 (H\textsc{ii}) \\ 
G208.80$-$02.65 & $2.8 \pm 0.7$ & $48.6 \pm 9.2$ & $0.69 \pm 0.17$ & S280 (H\textsc{ii}) \\ 
G234.20$-$00.20 & $8.2 \pm 0.7$ & $22.3 \pm 1.4$ & $0.68 \pm 0.11$ & S306, RCW 10 (H\textsc{ii}) \\ 

\bottomrule
\end{tabularx}
\end{table*}

\section{Global Sensitivity Metrics}
\label{Append: gsa concepts}

The analysis of sensitivity can also be interpreted as the evaluation of uncertainty of information. 
Therefore, before we introduce each metric for the global sensitivity analysis, we briefly introduce a few basic concepts in the probability theory and information theory.

\paragraph*{Entropy.} 
Entropy is a central concept in information theory that quantifies the average uncertainty, or equivalently the average information content, associated with the possible states of a random variable.
For a discrete random variable X taking values $x \in \mathcal{X}$ with probability mass function $p(x)\in[0,1]$, the entropy is defined as
\begin{equation}
    H(X) \defeq - \sum_{x\in\mathcal{X}} p(x) \log p(x).
\end{equation}
For continuous variables, the summation is replaced by an integral.
The definition of entropy for a single variable $X$ can be naturally generalised to the entropy of jointly distributed random variables. 
For example, the joint entropy of two variables, $X$ and $Y$, is
\begin{equation}
    H(X, Y) \defeq - \sum_{x\in\mathcal{X}, y\in\mathcal{Y}} p(x, y) \log p(x, y).
\end{equation}

\paragraph*{Conditional entropy.}
Conditional entropy quantifies the amount of information required to describe the outcome of a variable $Y$ given the known value of another variable $X$. 
It is given by
\begin{equation}
    H(Y|X) \defeq - \sum_{x\in\mathcal{X}, y\in\mathcal{Y}} p(x,y) \log \frac{p(x,y)}{p(x)},
\end{equation}
which satisfies
\begin{equation}
    H(Y|X) \defeq  H(X, Y) - H(X).
\end{equation}
More intuitively, conditioning on a variable $X$ removes the information contained in $X$.


\subsection{Global Sensitivity Metrics}
\label{sec:gsa}

\subsubsection{Mutual Information}
Mutual information (MI) is a measure of how much information variables $X$ and $Y$ share. In other words, it quantifies how much knowing one variable reduces uncertainty about the other. 
Mutual information can be conveniently defined as the difference between marginalised and joint entropies:
\begin{equation}
    I(X; Y) \defeq H(X) + H(Y) - H(X,Y) .
\end{equation}
It is always non-negative and equals zero if and only if $X$ and $Y$ are completely independent, i.e. $p(x,y)=p(x)p(y)$.

\subsubsection{Distance correlation}

Distance correlation (dCor) \citep{szekely2007measuring} is a measure that detects both linear and nonlinear associations between two random variables. 
This contrasts with Pearson's correlation \citep{pearson1895vii}, which can only detect linear associations.
Distance correlation is defined using the distance covariance between $X$ and $Y$, normalised by the product of their marginal distance variances. The distance correlation value ranges from $0$ (if and only if X and Y are independent) to $1$ (for perfectly dependent relationships).

\balance

\subsubsection{Permutation importance}
Permutation importance is a model-agnostic method for quantifying the contribution of each parameter to a predictive model. It evaluates how much the model’s performance degrades when the values of a parameter are randomly permuted, effectively measuring the sensitivity of the model to that parameter.
Formally, if $\mathcal{L}(\hat{O}, O)$ denotes the model's loss function, the permutation importance of parameter $X_j$ is
$I_j = \mathcal{L}(\hat{O}{\text{perm}(X_j)}, O) - \mathcal{L}(\hat{O}, O)$,
where $\hat{O}{\text{perm}(X_j)}$ is the model prediction after permuting $X_j$.

In this paper, we adopt a simplified approach to evaluating loss after permuting $X_j$, comparing the shuffled and unshuffled samples., i.e.
$I_j = \mathcal{L}(\hat{O}_{\text{perm}(X_j)}, O)$.
A larger value of $I_j$ indicates that the corresponding parameter has a greater impact on the model variation.
To summarise the results of permutations across multiple runs, we use \textbf{PermMean} to capture the average effect and \textbf{PermStd} to quantify the variability or uncertainty in the importance estimates. PermMean is the mean importance across permutations.

\subsubsection{Surrogate Gaussian Processes}

Surrogate modelling using Gaussian processes (GPs) is a widely used approach for approximating complex or costly functions. 
A GP defines a distribution over functions, providing both a predictive mean and an estimate of uncertainty at untested points. 

Global sensitivity analysis can be performed using the GP surrogate combined with Sobol indices. The first-order Sobol index $S_1^j$ measures the fraction of output variance attributable to a single input $X_j$. The total-order index $S_T^j$ quantifies the contribution of $X_j$. This includes all interactions with other inputs. Using a GP surrogate enables the efficient estimation of $S_1$ and $S_T$, even when the true model is costly to evaluate.
We use the confidence intervals, denoted $S_1^{\rm conf}$ and $S_T^{\rm conf}$, to quantify the uncertainty in the estimated first-order and total-order Sobol indices. These are usually obtained using either Monte Carlo resampling or GP surrogate propagation, which provides a measure of the statistical reliability of each sensitivity estimate. Note that Sobol indices assume that the inputs are independent over the sampled ranges. When the inputs are correlated, greater weight should be given to the model-free measures (MI and dCor) and to permutation importance.

When Automatic Relevance Determination (ARD) kernels are used, each input dimension $X_j$ is assigned a length scale, $\ell_j$, which quantifies how sensitive the function is to changes along that dimension. 
For convenience, one often considers $1/\ell_j$ as a measure of relative importance, with larger values indicating stronger dependence of the output on that parameter.

\subsubsection{Consensus Ranking}
To create a consensus ranking, we employ the aggregation-based (AggRank) method, which combines various feature importance metrics to generate a unified ranking. AggRank averages different importance measures, such as mutual information, permutation importance, ARD-based relevance and Sobol indices, to produce a robust overall ranking of parameters. This reduces sensitivity to the choice of a single metric and highlights the most influential variables across complementary analyses.



\section{Global Sensitivity Analysis of Distribution Parameters}

This appendix presents the global sensitivity analysis of the distribution parameters for the two SED summary features. The full set of results is summarised in Table~\ref{tab: gsa for ensemble model}. 

\begin{table*}
    \centering
    \caption{Global sensitivity analysis of the distribution parameters for the two SED summary features. Each block reports the same set of diagnostics for the listed parameters. The definitions and methodological details of these metrics are provided in Appendix~\ref{Append: gsa concepts}.}
    \label{tab: gsa for ensemble model}
    \begin{tabular}{lccccccccccc}
    \toprule
    Feature & Parameter & MI & dCor & PermMean & PermStd &  $S_1$ & $S_1^{\rm conf}$ & $S_T$ & $S_T^{\rm conf}$ & 1/ARD\_LS & AggRank \\
    \midrule
    \multirow{7}{*}{$\nu_{\rm p}$ (MC)} & $\lambda$ & 0.175 & 0.245 & 0.670 & 0.040 & 0.091 & 0.079 & 0.415 & 0.076 & 1.283 & 2.000 \\
     & $\ln{p_0}$ & 0.096 & 0.300 & 0.749 & 0.066 & 0.096 & 0.066 & 0.346 & 0.074 & 0.568 & 2.600 \\
     & $\gamma$ & 1.009 & 0.722 & 0.572 & 0.028 & 0.236 & 0.060 & 0.253 & 0.039 & 0.219 & 3.000 \\
     & $\sigma$ & 0.100 & 0.072 & 0.170 & 0.016 & 0.030 & 0.097 & 0.306 & 0.096 & 2.381 & 3.200 \\
     & $\ln a_0$ & 0.021 & 0.085 & 0.112 & 0.032 & 0.065 & 0.058 & 0.258 & 0.081 & 0.675 & 4.400 \\
     & $\delta$ & 0.024 & 0.045 & 0.008 & 0.001 & -0.003 & 0.006 & 0.005 & 0.001 & 0.107 & 5.800 \\
     & $\ln \Tilde{\beta}_0$ & 0.000 & 0.032 & 0.002 & 0.001 & 0.004 & 0.003 & 0.002 & 0.000 & 0.038 & 7.000 \\
    \midrule
    \multirow{7}{*}{$W$ (MC)} & $\gamma$ & 1.276 & 0.813 & 1.065 & 0.058 & 0.523 & 0.098 & 0.674 & 0.083 & 0.875 & 1.400 \\
     & $\lambda$ & 0.181 & 0.273 & 0.292 & 0.021 & 0.171 & 0.062 & 0.304 & 0.059 & 2.070 & 2.000 \\
     & $\sigma$ & 0.029 & 0.058 & 0.069 & 0.010 & -0.020 & 0.067 & 0.263 & 0.068 & 3.245 & 3.000 \\
     & $\ln a_0$ & 0.026 & 0.072 & 0.035 & 0.011 & 0.028 & 0.041 & 0.172 & 0.070 & 0.694 & 4.200 \\
     & $\ln{p_0}$ & 0.025 & 0.076 & 0.033 & 0.009 & 0.018 & 0.022 & 0.088 & 0.031 & 0.511 & 4.800 \\
     & $\delta$ & 0.013 & 0.050 & 0.006 & 0.001 & -0.003 & 0.026 & 0.085 & 0.049 & 0.863 & 5.800 \\
     & $\ln \Tilde{\beta}_0$ & 0.020 & 0.040 & 0.001 & 0.001 & -0.000 & 0.003 & 0.002 & 0.001 & 0.068 & 6.800 \\
    \midrule
    \multirow{7}{*}{$\nu_{\rm p}$ (DC)} & $\sigma$ & 0.102 & 0.131 & 0.779 & 0.089 & 0.114 & 0.222 & 0.747 & 0.153 & 7.818 & 2.000 \\
     & $\lambda$ & 0.108 & 0.183 & 0.338 & 0.029 & 0.025 & 0.040 & 0.191 & 0.041 & 1.458 & 2.800 \\
     & $\gamma$ & 1.424 & 0.746 & 0.520 & 0.039 & 0.118 & 0.046 & 0.162 & 0.040 & 0.422 & 3.000 \\
     & $\ln a_0$ & 0.019 & 0.115 & 0.742 & 0.177 & 0.053 & 0.072 & 0.589 & 0.212 & 1.164 & 3.600 \\
     & $\ln{p_0}$ & 0.050 & 0.203 & 0.260 & 0.027 & 0.013 & 0.028 & 0.146 & 0.035 & 0.539 & 4.200 \\
     & $\delta$ & 0.038 & 0.077 & 0.042 & 0.005 & -0.008 & 0.021 & 0.050 & 0.023 & 0.577 & 5.400 \\
     & $\ln \Tilde{\beta}_0$ & 0.002 & 0.048 & 0.021 & 0.006 & 0.009 & 0.012 & 0.010 & 0.003 & 0.165 & 7.000 \\
    \midrule
    \multirow{7}{*}{$W$ (DC)} & $\sigma$ & 0.129 & 0.116 & 0.755 & 0.132 & -0.030 & 0.059 & 0.568 & 0.382 & 27.787 & 1.600 \\
     & $\gamma$ & 1.697 & 0.870 & 0.794 & 0.127 & 0.241 & 0.135 & 0.398 & 0.333 & 0.831 & 1.800 \\
     & $\ln a_0$ & 0.035 & 0.116 & 0.545 & 0.207 & -0.012 & 0.039 & 0.437 & 0.274 & 1.744 & 2.600 \\
     & $\ln{p_0}$ & 0.031 & 0.074 & 0.011 & 0.003 & -0.000 & 0.008 & 0.005 & 0.003 & 0.141 & 4.400 \\
     & $\lambda$ & 0.002 & 0.041 & 0.018 & 0.009 & 0.001 & 0.006 & 0.006 & 0.008 & 0.118 & 5.000 \\
     & $\delta$ & 0.029 & 0.032 & 0.007 & 0.006 & 0.000 & 0.001 & 0.000 & 0.000 & 0.023 & 6.200 \\
     & $\ln \Tilde{\beta}_0$ & 0.004 & 0.027 & -0.001 & 0.006 & -0.000 & 0.001 & 0.000 & 0.000 & 0.046 & 6.400 \\
    \midrule
    \multirow{7}{*}{$\nu_{\rm p}$ (H~\textsc{ii})} & $\lambda$ & 0.382 & 0.455 & 1.346 & 0.071 & 0.264 & 0.109 & 0.789 & 0.134 & 2.859 & 1.000 \\
     & $\ln{p_0}$ & 0.378 & 0.447 & 1.103 & 0.100 & 0.169 & 0.086 & 0.660 & 0.174 & 1.132 & 2.200 \\
     & $\sigma$ & 0.078 & 0.052 & 0.031 & 0.004 & 0.025 & 0.049 & 0.079 & 0.033 & 1.522 & 3.600 \\
     & $\gamma$ & 0.153 & 0.159 & 0.019 & 0.003 & 0.010 & 0.011 & 0.017 & 0.006 & 0.180 & 4.400 \\
     & $\ln a_0$ & 0.005 & 0.048 & 0.020 & 0.005 & 0.033 & 0.056 & 0.089 & 0.071 & 0.723 & 4.800 \\
     & $\delta$ & 0.042 & 0.052 & 0.001 & 0.002 & 0.003 & 0.008 & 0.016 & 0.007 & 0.275 & 5.200 \\
     & $\ln \Tilde{\beta}_0$ & 0.005 & 0.043 & -0.002 & 0.003 & 0.001 & 0.005 & 0.004 & 0.002 & 0.091 & 6.800 \\
    \midrule
    \multirow{7}{*}{$W$ (H~\textsc{ii})} & $\lambda$ & 0.258 & 0.270 & 0.669 & 0.037 & 0.110 & 0.078 & 0.522 & 0.069 & 3.431 & 1.800 \\
     & $\ln{p_0}$ & 0.183 & 0.481 & 0.808 & 0.053 & 0.194 & 0.067 & 0.518 & 0.081 & 1.834 & 2.000 \\
     & $\gamma$ & 0.272 & 0.499 & 0.394 & 0.028 & 0.169 & 0.064 & 0.269 & 0.034 & 0.567 & 2.600 \\
     & $\sigma$ & 0.074 & 0.066 & 0.081 & 0.011 & -0.013 & 0.049 & 0.162 & 0.035 & 1.796 & 4.000 \\
     & $\ln a_0$ & 0.010 & 0.079 & 0.049 & 0.013 & -0.008 & 0.045 & 0.157 & 0.060 & 0.830 & 4.800 \\
     & $\ln \Tilde{\beta}_0$ & 0.018 & 0.052 & 0.003 & 0.002 & -0.003 & 0.009 & 0.010 & 0.003 & 0.157 & 6.400 \\
     & $\delta$ & 0.001 & 0.045 & 0.014 & 0.002 & -0.006 & 0.013 & 0.025 & 0.006 & 0.311 & 6.400 \\
    \bottomrule
    \end{tabular}
\end{table*}

\bsp	
\label{lastpage}
\end{document}